\documentclass[twocolumn]{aastex62}

\usepackage{url}
\usepackage[section]{placeins}
\usepackage{graphicx,times}
\usepackage[caption=false]{subfig}
\usepackage{amsmath} 
\usepackage{amssymb}

\usepackage{xcolor}

\newcommand{\nii}{[N \textsc{ii}]}
\newcommand{\sii}{[S \textsc{ii}]}
\newcommand{\oi}{[O \textsc{i}]}
\newcommand{\oiii}{[O \textsc{iii}]}
\newcommand{\feii}{[Fe \textsc{ii}]}
\newcommand{\brg}{Br$\gamma$}

\newcommand{\pab}{Pa$\beta$}
\newcommand{\molhy}{H$_2$}
\newcommand{\ha}{H$\alpha$}
\newcommand{\hb}{H$\beta$}
\newcommand{\fexxv}{Fe \textsc{XXV}}

\def\ltsima{$\; \buildrel < \over \sim \;$}
\def\simlt{\lower.5ex\hbox{\ltsima}}
\def\gtsima{$\; \buildrel > \over \sim \;$}
\def\simgt{\lower.5ex\hbox{\gtsima}}

\accepted{October 9, 2021}

\submitjournal{ApJ}

\shorttitle{Shocked Filaments in NGC~6240}
\shortauthors{Medling et al.}

\begin{document}

\title{Tracing the Ionization Structure of the Shocked Filaments of NGC~6240}

\correspondingauthor{Anne M. Medling}
\email{anne.medling@utoledo.edu}

\author[0000-0001-7421-2944]{Anne M. Medling}
\altaffiliation{Hubble Fellow}
\affil{Ritter Astrophysical Research Center, University of Toledo, Toledo, OH 43606, USA}
\affil{Research School of Astronomy \& Astrophysics, Mount Stromlo Observatory, Australia National University, Cotter Road, Weston, ACT 2611, Australia}
\affil{Cahill Center for Astronomy \& Astrophysics, California Institute of Technology, 1200 E. California Blvd., Pasadena, CA 91125, USA}
\affil{ARC Centre of Excellence for All Sky Astrophysics in 3 Dimensions (ASTRO 3D)}

\author[0000-0001-8152-3943]{Lisa J. Kewley}
\affil{Research School of Astronomy \& Astrophysics, Mount Stromlo Observatory, Australia National University, Cotter Road, Weston, ACT 2611, Australia}
\affil{ARC Centre of Excellence for All Sky Astrophysics in 3 Dimensions (ASTRO 3D)}

\author[0000-0002-5189-8004]{Daniela Calzetti}
\affil{Department of Astronomy, University of Massachusetts, Amherst, Amherst, MA 01003, USA}

\author[0000-0003-3474-1125]{George C. Privon}
\affil{National Radio Astronomy Observatory, 520 Edgemont Rd, Charlottesville,
VA, 22903, USA}
\affil{Department of Astronomy, University of Florida, 211 Bryant Space Sciences Center, Gainesville, FL 32611, USA}

\author[0000-0003-3917-6460]{Kirsten Larson}
\affil{Cahill Center for Astronomy \& Astrophysics, California Institute of Technology, 1200 E. California Blvd., Pasadena, CA 91125, USA}
\affil{Spitzer Science Center, California Institute of Technology, Pasadena, CA 91125, USA}

\author[0000-0002-5807-5078]{Jeffrey A. Rich}
\affil{The Observatories of the Carnegie Institution for Science, 813 Santa Barbara St., Pasadena, CA 91101, USA}

\author[0000-0003-3498-2973]{Lee Armus}
\affil{Spitzer Science Center, California Institute of Technology, Pasadena, CA 91125, USA}

\author[0000-0003-2168-0087]{Mark G. Allen}
\affil{Universit\'{e} de Strasbourg, CNRS, Observatoire astronomique de Strasbourg, UMR 7550, F-67000 Strasbourg, France}

\author[0000-0003-0234-7940]{Geoffrey V. Bicknell}
\affil{Research School of Astronomy \& Astrophysics, Mount Stromlo Observatory, Australia National University, Cotter Road, Weston, ACT 2611, Australia}

\author[0000-0003-0699-6083]{Tanio D\'{i}az-Santos}
\affil{N\'{u}cleo de Astronom\'{i}a de la Facultad de Ingenier\'{i}a y Ciencias, Universidad Diego Portales, Av. Ej\'{e}rcito Libertador 441, Santiago,
Chile}

\author[0000-0001-6670-6370]{Timothy M. Heckman}
\affil{Department of Physics \& Astronomy, Johns Hopkins University, Baltimore, MD, 21218, USA}

\author[0000-0003-2685-4488]{Claus Leitherer}
\affil{Space Telescope Science Institute, 3700 San Martin Dr, Baltimore, MD 21218}

\author[0000-0003-0682-5436]{Claire E. Max}
\affil{Department of Astronomy \& Astrophysics, University of California, Santa Cruz, CA 95064, USA}

\author[0000-0002-1608-7564]{David S. N. Rupke} 
\affil{Department of Physics, Rhodes College, Memphis, TN 38112, USA}

\author[0000-0001-7568-6412]{Ezequiel Treister}
\affil{Instituto de Astrof\'{i}sica, Facultad de F\'{i}sica, Pontificia Universidad Cat\'{o}lica de Chile, Casilla 306, Santiago 22, Chile}

\author[0000-0002-2985-7994]{Hugo Messias}
\affil{Joint ALMA Observatory and European Southern Observatory,
Alonso de C\'{o}rdova 3107, Casilla 19001, Vitacura, Santiago, Chile}

\author[0000-0002-5104-6434]{Alexander Y. Wagner}
\affil{Center for Computational Sciences, University of Tsukuba, 1-1-1 Tennodai, Tsukuba, Ibaraki 305-8577, Japan}

\begin{abstract}

We study the ionization and excitation structure of the interstellar medium in the late-stage gas-rich galaxy merger NGC~6240 using a suite of emission line maps at $\sim$25 pc resolution from the Hubble Space Telescope, Keck NIRC2 with Adaptive Optics, and ALMA.  NGC~6240 hosts a superwind driven by intense star formation and/or one or both of two active nuclei; the outflows produce bubbles and filaments seen in shock tracers from warm molecular gas (\molhy~2.12$\mu$m) to optical ionized gas (\oiii, \nii, \sii, \oi) and hot plasma (\fexxv).  
In the most distinct bubble, we see a clear shock front traced by high \oiii/\hb~and \oiii/\oi.  Cool molecular gas (CO(2-1)) is only present near the base of the bubble, towards the nuclei launching the outflow.  We interpret the lack of molecular gas outside the bubble to mean that the shock front is not responsible for dissociating molecular gas, and conclude that the molecular clouds are partly shielded and either entrained briefly in the outflow, or left undisturbed while the hot wind flows around them.
Elsewhere in the galaxy, shock-excited \molhy~extends at least $\sim$4 kpc from the nuclei, tracing molecular gas even warmer than that between the nuclei, where the two galaxies' interstellar media are colliding.  A ridgeline of high \oiii/\hb~emission along the eastern arm aligns with the south nucleus' stellar disk minor axis; optical integral field spectroscopy from WiFeS suggests this highly ionized gas is centered at systemic velocity and likely photoionized by direct line-of-sight to the south AGN.

\end{abstract}


\section{Introduction}

Local ultra- and luminous infrared galaxies (U/LIRGs, with log(L$_{\text{IR}}$/L$_{\sun})\geq$12 and 11, respectively) provide a useful probe of the energetic processes that shaped galaxies over cosmic time.  The high infrared luminosities of U/LIRGs are produced by dust heated by a combination of star formation and active galactic nucleus (AGN) activity \citep[e.g.][]{Sanders96}.  In the local Universe, these extreme infrared luminosities occur most commonly in gas-rich major mergers \citep{Veilleux02}, although mergers may not be required to trigger the same high star formation rates (SFRs) and AGN luminosities at z$\sim$2 \citep{Kartaltepe2010, Kartaltepe2012}.  To understand the evolutionary processes of galaxies at cosmic noon \citep[z$\sim$1-3, when star formation and AGN energy densities peaked;][]{madauplot,Aird10}, it is beneficial to study the detailed energetics of local galaxies with similar SFRs and AGN luminosities: U/LIRGs.

Emission line fluxes and ratios are powerful probes of the physical processes present in a galaxy.  Ratios of the optical emission lines \oiii/\hb, \nii/\ha, \sii/\ha, and \oi/\ha~can be used to distinguish between possible ionization mechanisms of gas using the BPT/VO87 line diagnostic diagrams \citep{BPT,Veilleux87}.  Although these diagnostic diagrams were originally used to differentiate between gas photoionized by HII regions and by AGN, recent models of fast \citep[$200-1000$ km s$^{-1}$;][]{Allen08} and slow \citep[$100-200$ km s$^{-1}$;][]{Rich10,Rich11} shocks have shown that the ionization mechanisms of galaxies are more complex.  For example, a galaxy containing star formation or AGN activity driving a wind may contain collisionally ionized shocked gas, increasing its line ratios into the AGN or LINER regions of diagnostic diagrams even if the photoionized gas shows lower line ratios.  For this reason, enhanced line ratios tracing shocks are now commonly associated with galactic winds \citep[e.g.][]{Ho14}.  Near-infrared emission lines such as \feii~and the ro-vibrational \molhy~transitions around 2.2 $\mu$m are also excited by outflows \citep{vanderWerf93,Veilleux09,Hill14}.  The interstellar media of galaxies must be clumpy and multiphase for a single wind to shock both molecular and ionized gas, because the shock velocities seen in ionized gas are fast enough to dissociate \molhy~\citep[$>$25 km s$^{-1}$;][]{Hollenbach80}.  Investigating shocks with multiple tracers is therefore required to fully understand the energy budget associated with galactic feedback.

NGC~6240 \citep[16h52m58.9s +02d24m03s, z = 0.0243, log(L$_{\text{IR}}$/L$_{\sun})=11.93$;][]{Kim13} is a particularly complex case study of shocks and shock drivers.  It hosts two X-ray confirmed AGN separated by 735 pc (1.5 arcsec) \citep{Komossa03,Max07}, is forming stars at up to 140 M$_{\sun}$ yr$^{-1}$ \citep{Heckman90}, and is driving a superwind seen in radio \citep{Baan07}, far-infrared \citep{Veilleux13}, near-infrared \citep{vanderWerf93, Max05}, optical \citep{Heckman90, Veilleux03}, and X-ray \citep{Nardini13, Wang14} observations.  Although NGC~6240's two nuclei both host AGN, the nuclei may be too weak or buried to dominate the ionization of the interstellar medium (ISM); star formation accounts for 76-80\% of the infrared luminosity \citep{Armus06}.  Star formation appears to be driving the outflow, which originates in the southern nucleus \citep{Baan07} and produces shock velocities up to the 2200 km s$^{-1}$ required to emit the diffuse hard X-ray photons \citep{Wang14}.  Roughly 3$\times$10$^{9}$ M$_{\sun}$ of molecular gas is present in a tidal bridge between the two nuclei \citep{Tacconi99,Engel10}.  The majority of molecular line emission is seen in this region and likely comes from C-shocks with $v_{shock}  = 20-40$ km s$^{-1}$ produced by the collision of the interstellar media of the two galaxy nuclei \citep{vanderWerf93}.  Shocked molecular gas is also present in the superwind, demonstrated by the filaments of \molhy~\citep{Max05} and CO emission \citep{Feruglio13a, Feruglio13b} tracing the ionized gas emission features.  \citet{Feruglio13b} estimates that 100-500 M$_{\sun}$ yr$^{-1}$ of molecular gas is expelled by this outflow. 

In this paper, we present high spatial resolution ($\sim$ 50 milliarcsecond = 24.5 parsec) maps of the inner few kiloparsecs of NGC~6240 highlighting key optical, near-infrared, and sub-millimeter emission lines, from narrowband imaging and 3D spectroscopy.  This unique combined dataset offers the most comprehensive view to date of the structure of the ISM in a galaxy hosting AGN, star formation, outflows, and shocks.
In Section~\ref{obs}, we describe our data and data processing techniques.  We compare the fluxes and morphologies of emission lines in Section~\ref{linestructure} and discuss their physical implications in Sections~\ref{bubbles},~\ref{dust}, and \ref{oiiiarm}.  
In Section~\ref{conclusions}, we summarize our conclusions.  Throughout the paper we adopt $H_0 = 70$\,km\,s$^{-1}$\,Mpc$^{-1}$,
$\Omega_{\rm m}$ = 0.28, and $\Omega_\Lambda$ = 0.72 \citep{Hinshaw09}.  The physical scale is thus 490 pc/arcsec, calculated using Ned Wright's Cosmology Calculator\footnote{Available at \url{http://www.astro.ucla.edu/~wright/CosmoCalc.html}.} \citep{Wright06}.


\section{Observations}
\label{obs}

\subsection{Imaging}
\subsubsection{Hubble Space Telescope Imaging}

We used a combination of narrow-, medium-, and broad-band optical images from the Hubble Space Telescope's Wide-Field Camera 3 (WFC3) and Advanced Camera for Surveys (ACS) to isolate key emission lines.  Most optical images were observed in Cycle 19 (Proposal ID: 12552; PI: L. Kewley); the F814W image was observed in Cycle 14 (Proposal ID: 10592, PI: A. Evans); images in the 1-2 $\mu$m range were observed in Cycle 22 (Proposal ID: 13690, PI: T. Diaz-Santos).  Details of these observations are listed in Table~\ref{tbl:obs}.  The fully-reduced and calibrated images from ACS and WFC3/UVIS were downloaded from the Barbara A. Mukulski Archive for Space Telescopes in January of 2019.  The WFC3/IR images (F130N and F132N) were specifically regridded and processed as described in Larson et al. (submitted) before our analysis.

\subsubsection{Keck/NIRC2 Adaptive Optics Imaging}

We also make use of Keck Adaptive Optics (AO) narrowband imaging with the NIRC2 camera (PI: K. Matthews) around 2.2 $\mu$m.  These images were initially presented in \citet{Max05}.  The Keck II AO system \citep{Wiz00, vanDam04} used a natural guide star (R=11.9, B=13.5, separation = 35.8\arcsec) and a deformable mirror to measure and correct for atmospheric turbulence and provide near diffraction-limited image resolution.  Isolated star clusters in the image show a full-width at half max of 0.14\arcsec, an upper limit on our achieved spatial resolution.  The imaging presented here uses the wide camera (0.04\arcsec pixel$^{-1}$), a good match to the ACS imaging plate scale (0.05\arcsec pixel$^{-1}$).  No observations of photometric standards were taken with NIRC2 during this run; to flux calibrate the data, we used archival NICMOS imaging of in-field star clusters in similar narrowband filters.

\subsubsection{Registering the Images}
We registered all images to match the F621M filter coordinates using the IRAF routines geomap and geotran.  We used the coordinates of isolated non-saturated stars (typically 15-25 per filter) as inputs to the geomap routine; this number is sufficient to determine the translational shift, rotation, and distortion corrections for the WFC3 camera.  

\begin{table}
\caption{Narrowband Imaging Emission Line Content}\label{tbl:obs}
\begin{tabular}{lcccc}
\hline
\hline
Filter & Exp. Time & Spectral Range & Included Strong & \\
 &  (s) & (nm) & Emission Lines \\
\hline
\hline
\textbf{HST WFC3} & \multicolumn{3}{l}{\textbf{Proposal ID:12552}} \\
F467M  & 1540 & 458.25-478.35 & -- \\
FQ492N  & 5490\tablenotemark{a} & 487.6-499.9 & \hb \\
FQ508N  & 5490\tablenotemark{a} & 502.55-515.65 & \oiii \\
F621M  & 310 &  591.45-652.35 &  \oi \\
F645N  & 2358 & 641.2-649.6 & \oi \\ 
F673N  & 390 & 670.7-682.5 & \ha, \nii \\
F680N  & 1050 & 669.19-706.25 & \ha, \nii, \sii \\
& \multicolumn{3}{l}{\textbf{Proposal ID: 13690}} \\
F130N & 5383 & 1292.8-1308.4 & --\\ 
F132N & 5383 & 1310.75-1326.85 &  \pab  \\
\hline
\textbf{HST ACS} & \multicolumn{3}{l}{\textbf{Proposal ID: 10592}} \\
F814W  & 720 & 773.1-838.3 & -- \\
\hline
\textbf{Keck NIRC2}   \\
\brg & 1200 & 2152.3-2184.9& \molhy \\
\molhy \tablenotemark{b}& 1200 & 2111.2-2145.2 & -- \\
\hline 

\end{tabular}
\tablenotetext{a}{Two additional frames totalling 2160 and 2090 seconds for each of FQ492N and FQ508N respectively exist but were excluded from the analysis due to excessive cosmic rays.}
\tablenotetext{b}{This filter was formerly incorrectly labeled `NB2108' in the vendor documentation.}
\end{table}

\subsubsection{Line Maps}

Careful continuum subtraction is critical to producing meaningful emission line maps from narrowband imaging, particularly in a galaxy with high dust extinction like NGC~6240 \citep[global A$_{V}\sim$ 15-20 mag;][]{Lutz03}.  We counteract this by producing clean continuum maps in three bracketing wavelength bands; the final continuum subtraction for each filter is calculated by interpolating the continuum spectrum smoothly between continuum maps.

\begin{figure*}
\centering
\includegraphics[width=\linewidth]{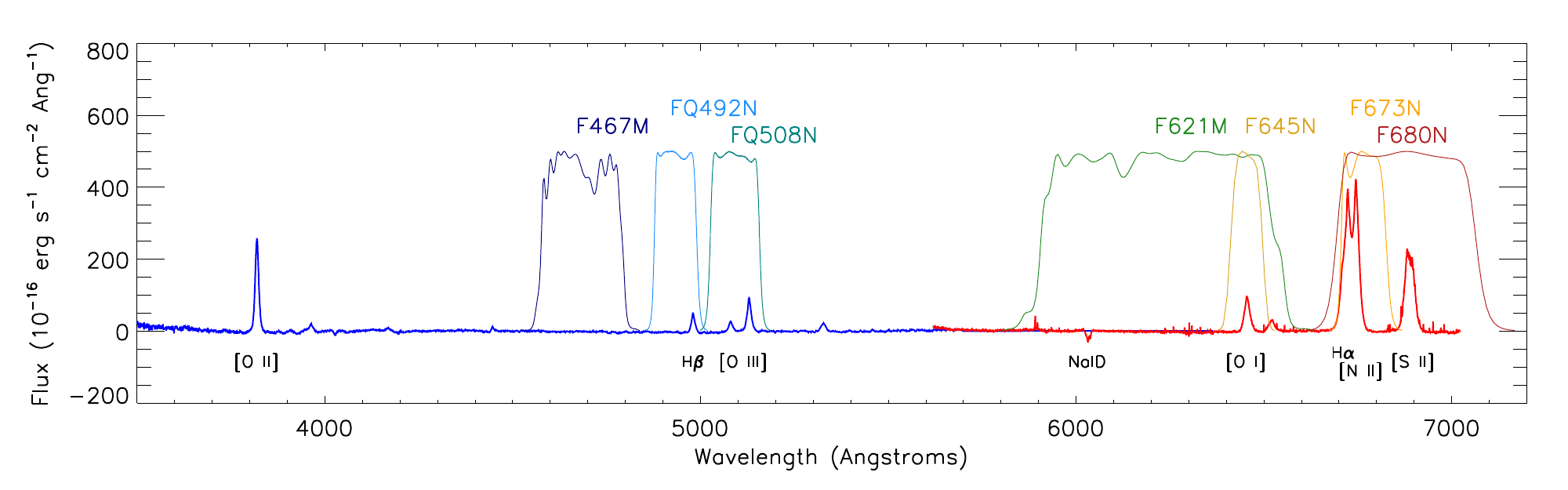}
\caption{Representative optical spectrum of NGC~6240 integrated over our WiFeS field-of-view showing the blue (wavelength $<$ 5800$\AA$) and red (wavelength $>$ 5700$\AA$) arms, with total HST system throughput curves of relevant WFC3/UVIS filters overlaid.}  
\label{filters}
\end{figure*}

Figure~\ref{filters} shows a representative optical spectrum of NGC~6240 from our WiFeS data (see Section~\ref{wifes}) with the relevant throughput curves for each HST filter overlaid.  The throughput curves are the most up-to-date version as of May 31, 2018\footnote{Throughput curves were downloaded from \url{http://www.stsci.edu/hst/wfc3/ins_performance/throughputs/Throughput_Tables}}.  The F467M filter is clean of strong emission lines; we use it as the blue continuum map.  The F814W filter is broad (65.2 nm wide) and contains relatively few emission lines.  We estimate the emission line contamination of the F814W is $<$1\%, 
and use it as the reddest continuum map.  We also use the combination of F621M and F645N images (which both contain only \oi) to create an intermediate continuum map for better interpolation, with the following iterative procedure: 
\begin{enumerate}
\item{With F621M and F645N images in erg s$^{-1}$ cm$^{-2}$ \AA$^{-1}$, the \oi~contribution to the medium-band filter is significantly lower.  We use it to provide a first estimate of the continuum and create an initial \oi~map. \\
$\text{map}_{\oi,0}$ = F645N - F621M
}
\item{Because this \oi~map estimate is in erg s$^{-1}$ cm$^{-2}$ \AA$^{-1}$, we scale it by the ratio of bandwidths and subtract it from the medium-band filter to obtain a cleaned continuum map. \\
$\text{map}_{cont,1}$ = F621M - $\text{map}_{\oi,0} \times \frac{\Delta \lambda_{\text{F645N}}}{\Delta \lambda_{\text{F621M}}}$
}
\item{With this improved continuum estimate, we produce a cleaned \oi~map. \\
$\text{map}_{\oi,1}$ = F645N - $\text{map}_{cont,1}$ }
\item{We iterate again to confirm the \oi~map is stable: \\
$\text{map}_{cont,2}$ = F621M - $\text{map}_{\oi,1}\times \frac{\Delta \lambda_{\text{F645N}}}{\Delta \lambda_{\text{F621M}}}$ \\
$\text{map}_{\oi,2}$ = F645N - $\text{map}_{cont,2}$ }

\item{The differences between the second and first iteration of $map_{\oi}$ is approximately 5\% of the image's 1-$\sigma$ uncertainty level, which we find sufficient.
Because $map_{\oi,2}$ $\approx$ $map_{\oi,1}$, we are confident in our continuum subtraction, and adopt $map_{cont,2}$ as a clean continuum map at intermediate wavelengths.  We multiply $map_{\oi,2}$ by the bandwidth of F645N to produce a final \oi~line map in units of erg s$^{-1}$ cm$^{-2}$.}

\end{enumerate}

To produce the \hb~and \oiii~line maps, we interpolate between the F467M clean continuum and the continuum from the cleaned F621M image to the appropriate wavelength.  Our final line maps have this continuum subtracted, and are then multiplied by the relevant bandwidth in order to recover the units of erg s$^{-1}$ cm$^{-2}$.  A map of \ha+\nii~is constructed in the same way, using a continuum interpolated between that of the F621M filter and the F814W filter.  To produce the \sii~line map (which for this paper always refers to the sum of the fluxes of the \sii~$\lambda$$\lambda$6716,6731 doublet), we subtract off both the continuum and the \ha+\nii~contribution from the F680N filter.  Note that Balmer emission lines are therefore \textit{not} corrected for stellar absorption.  Based on the spectroscopy described in Section~\ref{wifes}, we estimate the effect of stellar absorption at the $<$15\% level in \hb~across most of the galaxy; in a few percent of spaxels, the Balmer absorption correction can be as high as 50\% for \hb.  The deep stellar absorption therefore contributes to the low S/N seen in our \hb~map, particularly around the nuclei.  Maps of \hb~equivalent width (of the continuum and the overall data) from our integral field spectroscopy are shown in Appendix~\ref{wifesappendix}.

Both near-infrared line maps (\pab~and \molhy) come from a matched pair of filters: one containing the line and one containing nearby continuum.  As above, the line images have the continuum contribution removed and are then multiplied by the relevant bandwidth to present the line flux in units of erg s$^{-1}$ cm$^{-2}$.    Maps and contours of stellar continuum in figures below use the F130N image, because the reddest line-free image is least affected by dust.

Due to the presence of noise in the images, we mask out flux that is below the 3$\sigma$ level.  The noise level of each line map is given as the standard deviation of pixel values in an off-galaxy region 4\arcsec by 4\arcsec in size, 30\arcsec from the nucleus at 100$^{\circ}$ east of north.
Final line maps are shown in Figure~\ref{linemaps} and discussed in the following sections.

\subsection{Spatially Resolved Spectroscopy}

We also incorporate optical integral field spectroscopy (IFS) of NGC~6240 into our analysis, to estimate the \nii/\ha~ratios and to provide kinematic interpretations in later sections, Band 6 interferometry from the Atacama Large Millimetre Array (ALMA) to map the CO(2-1) cold gas emission, and Chandra X-ray observations to trace hot gas.  

\subsubsection{Integral Field Spectroscopy from the Wide-Field Spectrograph on the ANU 2.3-m Telescope}
\label{wifes}

We use observations of NGC~6240 from WiFeS on the ANU 2.3-m Telescope at Siding Spring Observatory, taken in July 2014, with 10-minute frames taken at a ratio of approximately three target frames to one sky frame, totaling 120 minutes on source.  WiFeS \citep{Dopita07,Dopita10} is a dual-arm image slicer that, in this mode, covers 3700-5700\AA~at a spectral resolution of R$\sim$3000 and 5700-7000\AA~at R$\sim$7000; WiFeS frames cover an area of 25\arcsec$\times$36\arcsec, with 1\arcsec~spatial sampling.  During these observing runs, the seeing was approximately 1\farcs5.

Our WiFeS data were reduced using the standard PyWiFeS pipeline v0.6.0 \citep{Childress14}.  The resulting blue and red data cubes were median-combined and then analyzed using LZIFU \citep{LZIFU}, an IDL package that uses penalized pixel fitting \citep[pPXF;][]{PPXF} to fit the stellar continuum and then MPFIT \citep{mpfit} to fit one, two, or three Gaussian components to each of 11 strong emission lines simultaneously.  Across most of the field of view, the line profiles are extremely complex, and a standard f-test recommends using all three Gaussian components.  We therefore estimate the \nii/\ha~ratio in Section~\ref{linestructure} using the sum of all three components for each line.

Select lines and line ratios are presented in the main body of the text where higher spatial resolution HST data are unavailable, but the entire set of emission line maps is presented in Appendix~\ref{wifesappendix} for completeness.

\subsubsection{Band 6 Interferometry from ALMA}

We present new long-baseline imaging of CO(2-1) from the Atacama Large Millimetre Array (ALMA).  This Band 6 observation was taken as part of program 2015.1.00370.S (PI: Treister) in three blocks between November 2015 and May 2016 for an aggregate exposure time of approximately 4600 seconds and has a resolution of 0.06\arcsec $\times$ 0.03\arcsec.  To improve the coverage of larger-scale structure, the data from program 2015.1.00003.S \citep[PI: Saito;][]{Saito18} were also included.  

These data provide the highest resolution view of the cold molecular gas to date, which we contrast with other phases of the ISM at similar spatial resolutions in this paper.  A complete analysis of the CO(2-1) and Band 6 continuum dataset is presented in \citet{Treister20}.

\subsubsection{Chandra X-ray Observatory Data}

We make use of deep X-ray observations from the \textit{Chandra X-ray Observatory}, originally published in \citet{Nardini13} and \citet{Wang14} and graciously provided by those authors in their processed forms.  The observations are comprised of $\sim$150 ks of integration using the \textit{Chandra} Advanced CCD Imaging Spectrometer \citep[ACIS;][]{Garmire03}.
Here we present the hard X-ray continuum (5.5-8 keV) and \fexxv emission line maps from \citet{Wang14} and the soft X-ray continuum (0.7-1.1 keV) map from \citet{Nardini13}.  All three are tracers of hot gas likely associated with winds.

\section{Results: Emission Lines and Ratios}
\label{linestructure}

\subsection{Emission Line Structure}
	
\begin{figure*}
\centering
\includegraphics[width=\linewidth,trim=1.25cm 0.5cm 0cm 1cm,clip]{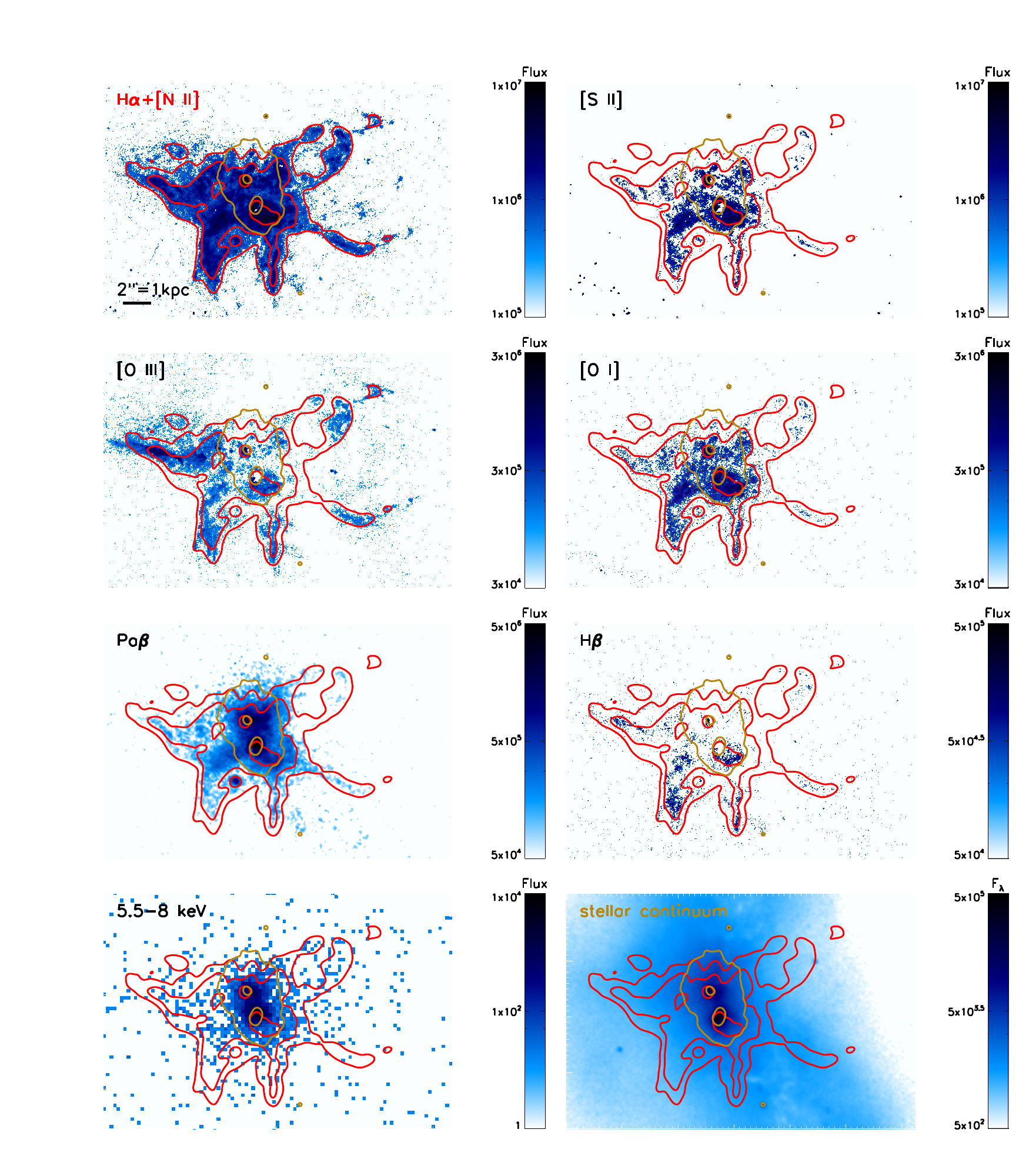}
\caption{Maps of emission lines \ha+\nii, \sii, \oiii, \oi, \pab, \hb, \molhy 1-0 S(1), CO(2-1), \fexxv, hard and soft X-rays, and the stellar continuum from the F130N WFC3 filter.  Each panel is shown in log scale with limits chosen to emphasize structure, and has \ha+\nii~contours (red) and stellar continuum~contours (brown) overlaid for comparison.   Optical/NIR line emission flux maps are shown in units of 10$^{-20}$ erg s$^{-1}$ cm$^{-2}$ arcsec$^{-2}$ and the stellar continuum flux map in units of 10$^{-20}$ erg s$^{-1}$ cm$^{-2}$ \AA$^{-1}$ arcsec$^{-2}$.  The X-ray maps are shown in units of counts arcsec$^{-2}$ and the CO(2-1) map in units of Jy beam$^{-1}$ km s$^{-1}$.  The scale bar in the top left panel shows 2.05 arcseconds, approximately one kiloparsec.
}  
\label{linemaps}
\end{figure*}

\renewcommand{\thefigure}{\arabic{figure} (Cont.)}
\addtocounter{figure}{-1}

\begin{figure*}
\centering
\includegraphics[width=\linewidth,trim={1.25cm 10cm 0 1cm},clip]{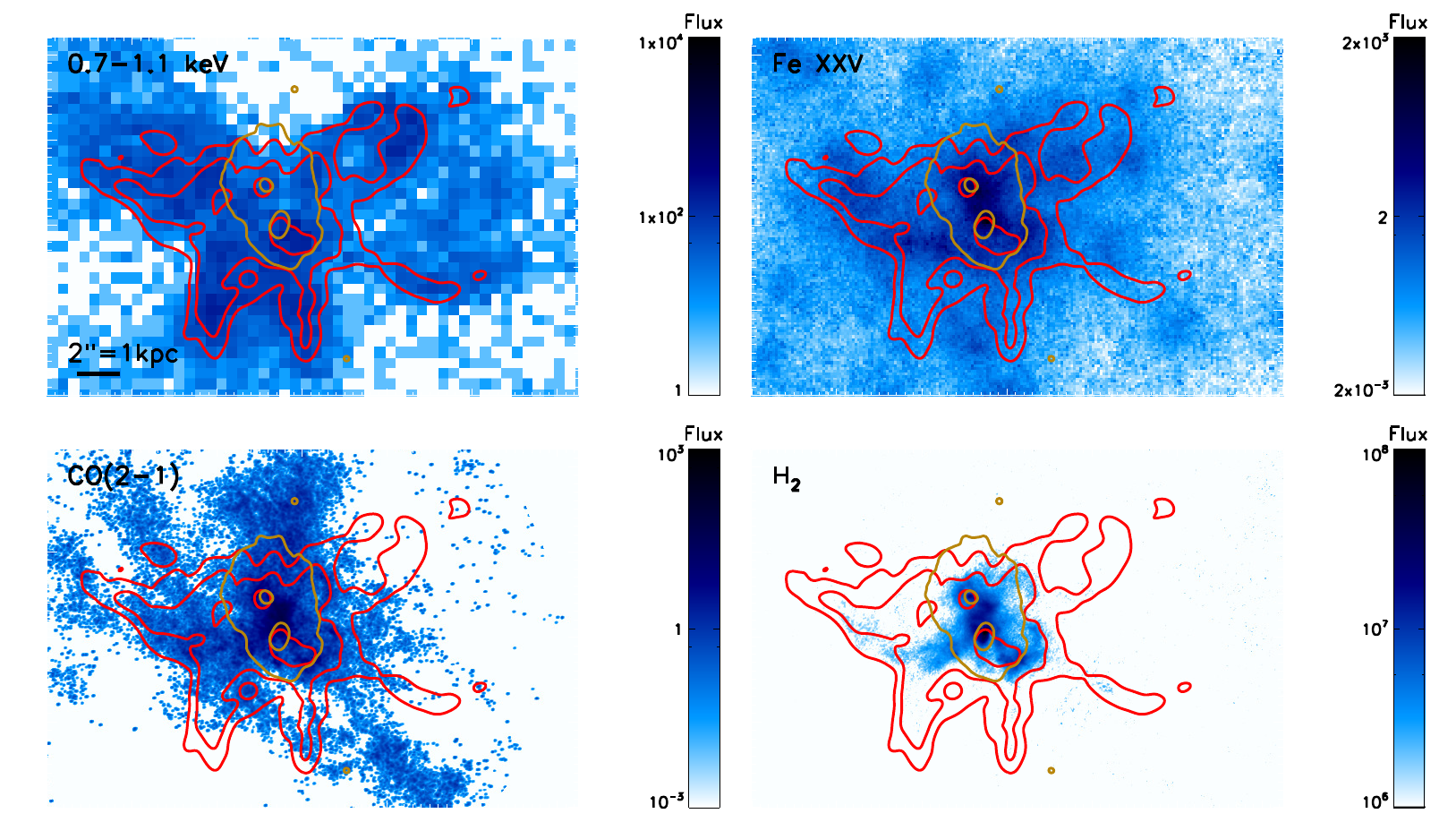}
\caption{Maps of emission lines \ha+\nii, \sii, \oiii, \oi, \pab, \hb, \molhy 1-0 S(1) and the stellar continuum from the F130N WFC3 filter.  Each panel is shown in log scale with limits chosen to emphasize structure, and has \ha+\nii~contours (red) and stellar continuum~contours (brown) overlaid for comparison.   Optical/NIR line emission flux maps are shown in units of 10$^{-20}$ erg s$^{-1}$ cm$^{-2}$ arcsec$^{-2}$ and the stellar continuum flux map in units of 10$^{-20}$ erg s$^{-1}$ cm$^{-2}$ \AA$^{-1}$ arcsec$^{-2}$.  The X-ray maps are shown in units of counts arcsec$^{-2}$ and the CO(2-1) map in units of Jy beam$^{-1}$ km s$^{-1}$. The scale bar in the top left panel shows 2.05 arcseconds, approximately one kiloparsec.
}  
\end{figure*}
\renewcommand{\thefigure}{\arabic{figure}}

\begin{figure}
\centering
\includegraphics[scale=.32,trim=1.25cm 0.5cm 0cm 1cm,clip,angle=-90]{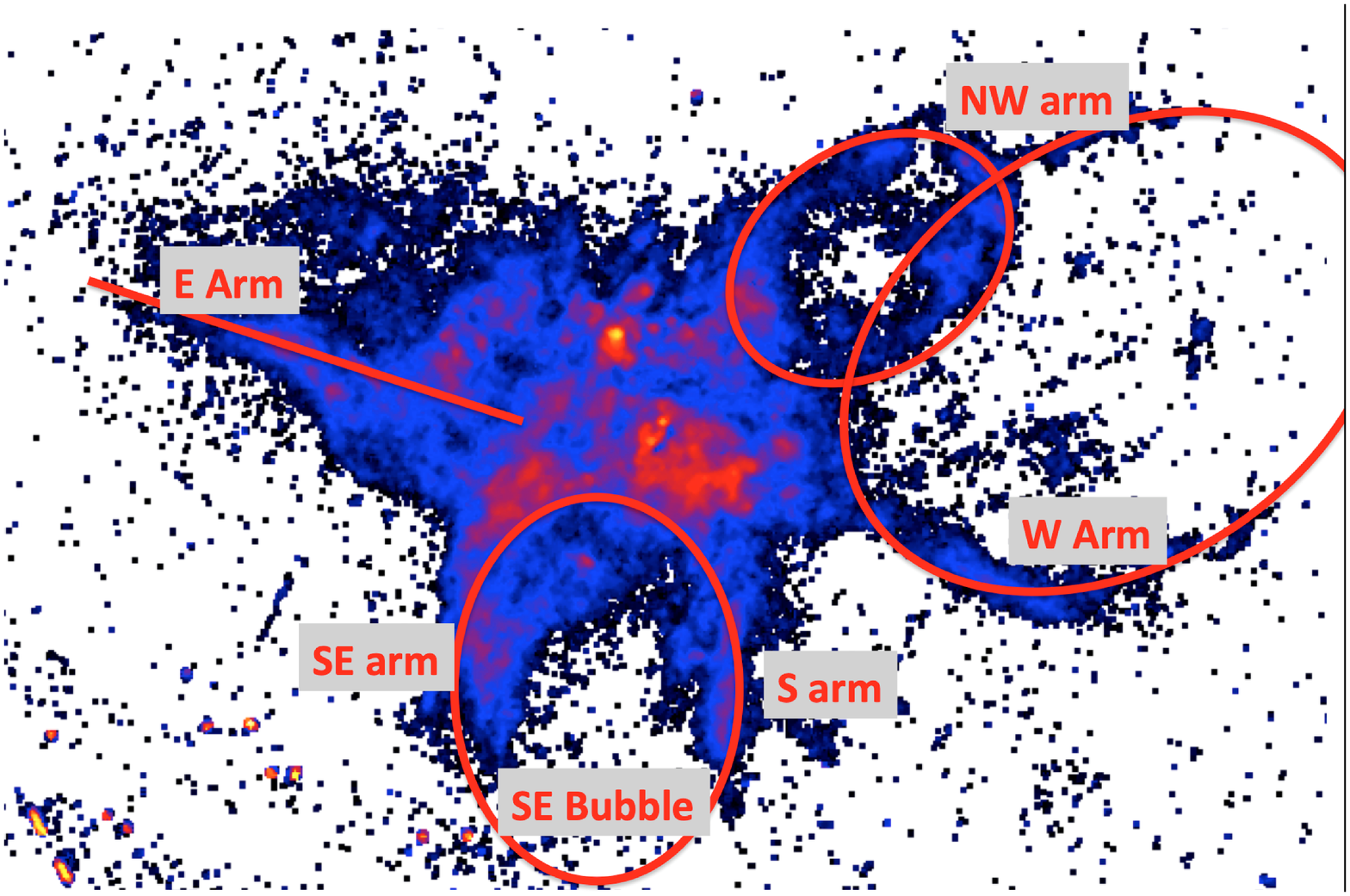}
\caption{Map of \ha+\nii~from Figure~\ref{linemaps} with schematic identification of the filamentary and bubble features referred to throughout this work.
} 
\label{bubblesketch}
\end{figure}

In Figure~\ref{linemaps}, we present flux maps of the key emission lines \ha+\nii, \sii, \oiii, \oi, \pab, \hb, and the near-infrared stellar continuum from HST, hard and soft X-ray continuum and the \fexxv~line from Chandra, warm \molhy~emission from Keck, and CO(2-1) emission from ALMA.  All emission lines presented in Figure~\ref{linemaps} show filamentary and/or bubble structure associated with the superwinds and merger disruptions known to be present in NGC~6240.  The \ha+\nii~emission is the deepest, and shows similar features to those seen by \citet{Heckman87,Heckman90} and \citet{Keel90}; a major result of the starburst-driven superwind is the `hourglass' oriented at a position angle 100$^{\circ}$ east of north.  Throughout the text we refer to various bubbles and arms; for clarity we sketch them in Figure~\ref{bubblesketch}.

The \ha+\nii~emission has been studied extensively by \citet{Yoshida16}, who found a network of bubbles and filaments extending 70$\times$80 kpc, linked to three epochs of starburst-driven superwinds.  In general, the \ha+\nii~emission tracks the extended soft X-ray halo \citep{Nardini13}, suggesting that these gas phases are linked.  \citet{Max05} compared the central \ha+\nii~emission to \molhy~emission and found qualitative similarities; this correlation confirms that the superwind is affecting gas of a wide range of densities.  Indeed, most filamentary features in the \molhy~emission have matching features in the CO(2-1) and the \fexxv~maps \citep[the latter of which was pointed out in][]{Wang14}.  We examine the spatial structure of \molhy~in the bubbles in Section~\ref{bubbles}.

Although the \ha+\nii~emission has been well-studied, the complexity of the kinematic structure has prevented a detailed study of optical emission line ratios in the Butterfly Nebula \citep[e.g.][]{Sharp10}.  Our suite of optical and near-infrared emission line maps is the most complete to date.  Most of the morphological differences between the various optical/NIR lines are driven in large part by dust attenuation.  The blue lines \hb~and \oiii~ are heavily extincted in the nucleus, and as we probe longer wavelengths, the nuclear region becomes much brighter.  The most striking individual feature present is the intense, collimated \oiii~emission along the eastern arm.  Although emission in other lines is present in that region, they do not peak along the same spatial line.  

\begin{figure*}
\centering
\includegraphics[width=\linewidth,trim=0.2cm .5cm 0.2cm 1cm,clip]{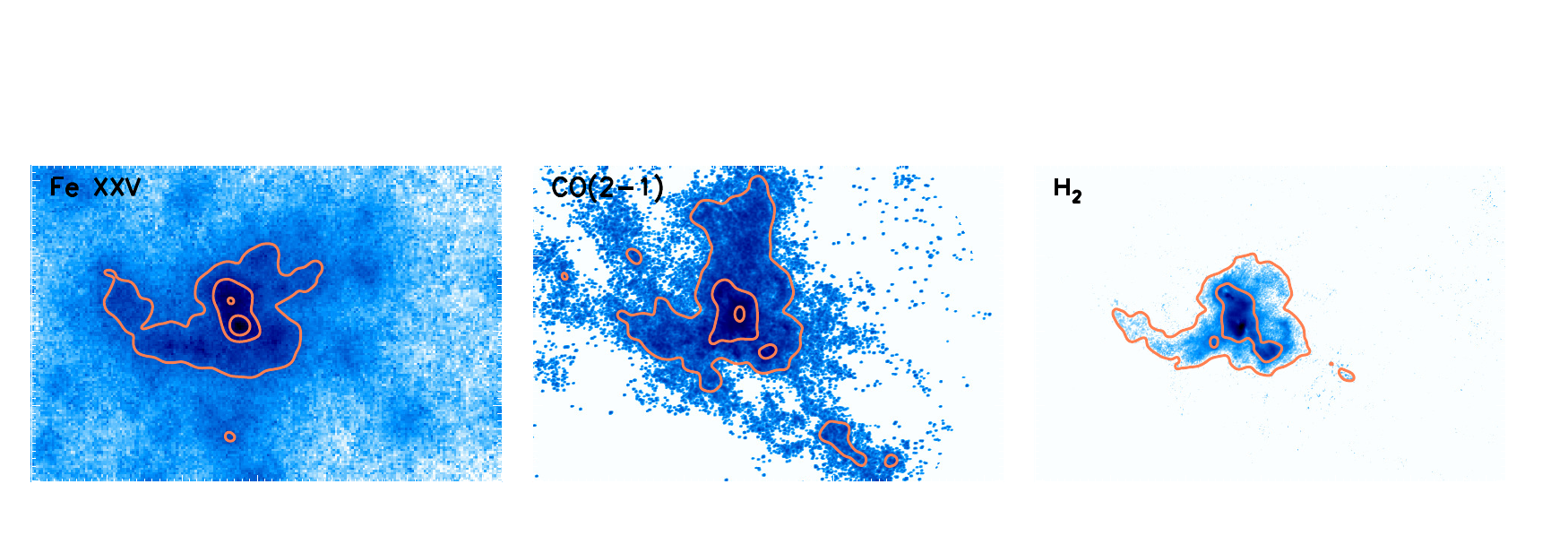}
\caption{
Maps of \fexxv, CO(2-1), and \molhy~emission as in Figure~\ref{linemaps}, but with individual contours overlaid in orange to highlight similar morphologies.  All three emission lines are bright in the nuclei and extend north, southwest, and (most prominently) $\sim$3 kpc to the southeast.
}  
\label{morphcompare}
\end{figure*}

As noted in \citet{Wang14}, the \fexxv~emission matches the \molhy~in overall morphology.  
To highlight the similarity, we reproduce several line maps from Figure~\ref{linemaps} with contours overlaid.  These maps shown in Figure~\ref{morphcompare} reinforce that 
where we detect \molhy~emission, we not only see \fexxv~but also CO(2-1), particularly in extension to the southeast.  Although some similarity may be only in projection, the presence of the same $\sim$3 kpc loop to the southeast suggests that these regions likely represent structures containing truly multiphase gas: ranging from $\sim$10K to 7$\times$10$^{7}$K.

\subsection{Emission Line Ratios}
\begin{figure*}
\centering
\includegraphics[width=\linewidth,trim=0 0 1cm 0,clip]{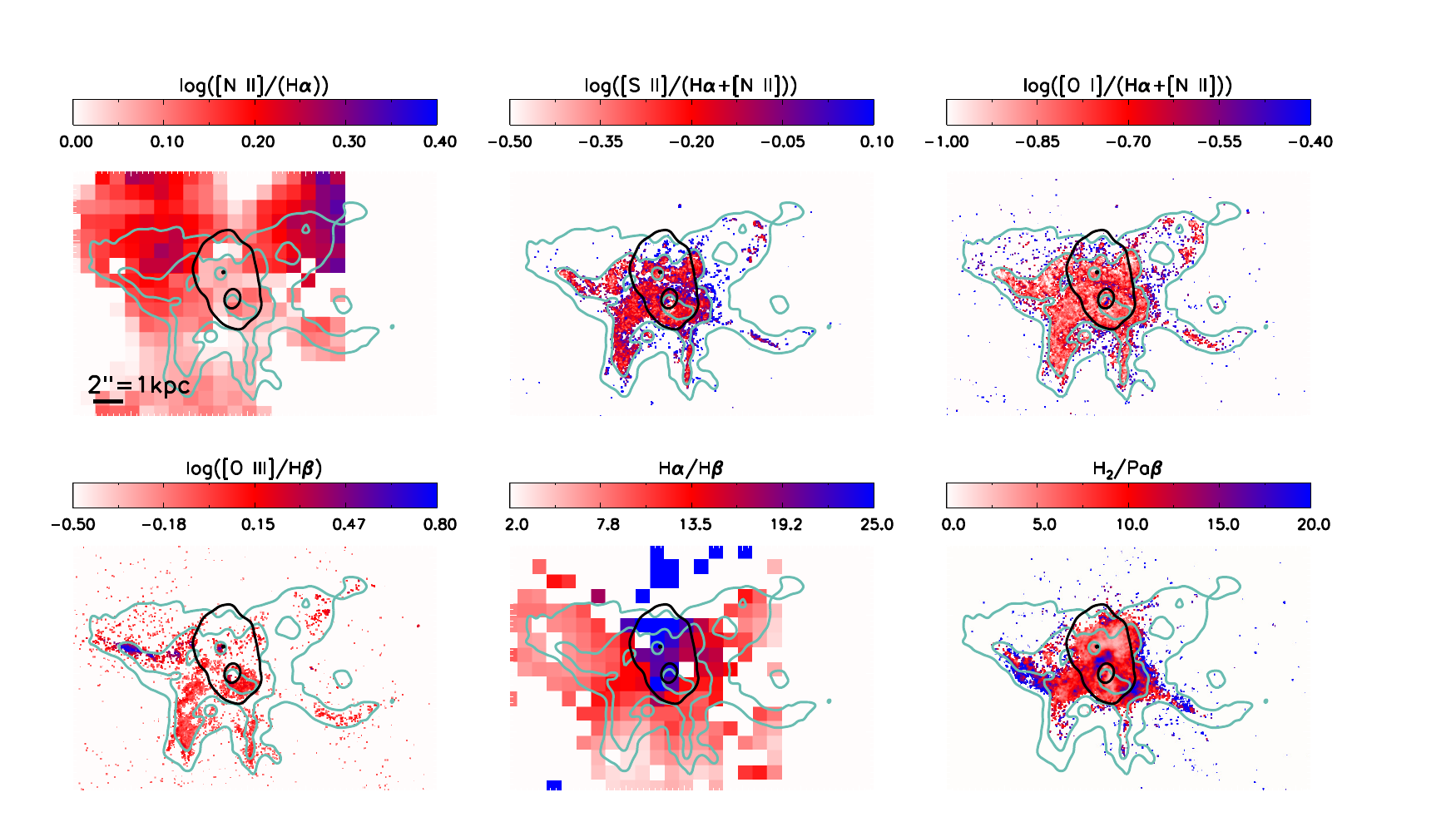}
\caption{Maps of emission line ratios (top row, left to right), log(\nii/\ha) from WiFeS, log(\sii/(\ha+\nii)), log(\oi/(\ha+\nii)), and (bottom row, left to right) log(\oiii/\hb), \ha/\hb~from WiFeS, and \molhy 1-0 S(1)/\pab.  HST line maps are initially masked below 3$\sigma$ and then median-filtered with a kernel of 3x3 pixels.
Each panel has the \ha+\nii~contours (teal) and stellar continuum~contours (black) from Figure~\ref{linemaps} overlaid for comparison.  The scale bar in the top left panel shows 2.05 arcseconds, approximately one kiloparsec.
}  
\label{ratiomaps}
\end{figure*}

Emission line maps are complex to interpret because they show a convolution of intrinsic gas properties like density and metallicity with extrinsic properties like the radiation field or shocks that are currently ionizing the gas.  Further, dust extinction will affect how much of the flux emitted in each line reaches our telescopes, in a spatially-varying way.  Some of these degeneracies can be limited by looking at emission line ratios.  For example, pairs of emission lines close in wavelength will be affected by dust to a similar degree, and photoionization or shock models can predict the ratio of those two lines given particular gas properties and conditions.  Line ratios are more exact than attempting to correct for extinction, which is challenging when different physical components (even along a single line-of-sight) are enshrouded by varying amounts of dust.  

In Figure~\ref{ratiomaps}, we show the emission line ratio maps constructed from the line maps described in Figure~\ref{linemaps} smoothed by a median filter with a 3x3 pixel kernel to emphasize larger-scale structure, with two additions: the log(\nii/\ha) and the Balmer decrement \ha/\hb~maps from our ground-based WiFeS integral field spectroscopy.  We use the Balmer decrement from WiFeS here to show extinction instead of the HST-based \pab/\hb~ratio because the HST \hb~image is not deep enough to probe the heavily-extincted nuclear regions.

\begin{figure*}
\centering
\includegraphics[width=\linewidth,trim=0 0 0cm 0,clip]{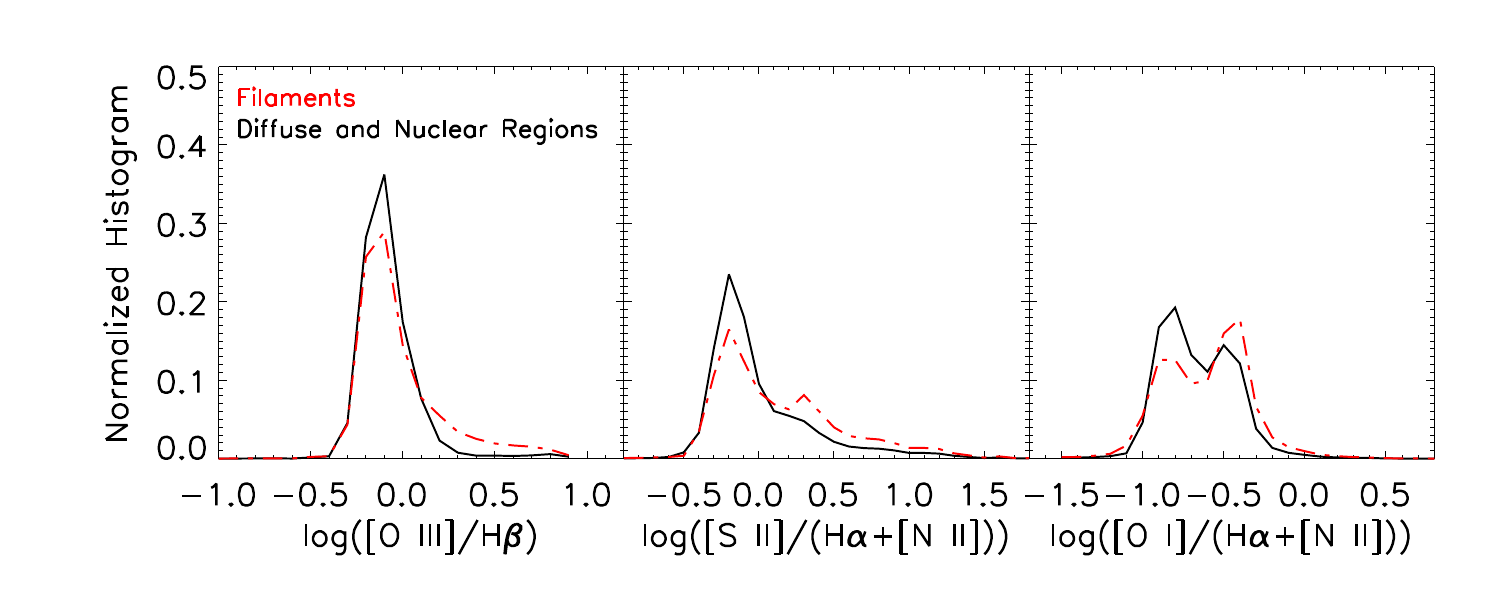}
\caption{Normalized histograms showing the distribution of line ratios present in HST narrowband maps.  In each panel, the red dot-dashed line shows the distributions present along the filaments and edges of bubbles marked in Figure~\ref{bubblesketch} and the black solid line shows the remaining spatial regions of the galaxy.  In each panel, the filamentary regions have a tail towards higher line ratios.  A K-S test confirms that the two distributions are drawn from different populations (p$<10^{-7}$) in each case.
}  
\label{filament_histograms}
\end{figure*}

The line ratio maps show several interesting features of the filamentary structure. 
The distributions of line ratios show a statistically significant excess of higher values in the filaments, compared to the diffuse and nuclear regions (Figure~\ref{filament_histograms}).
Across the three optical line ratios probed, filamentary regions show a significant tail to higher ratios.  A two-sided Kolmogorov-Smirnov test shows that the two samples are drawn from different populations to a strong significance (p$<10^{-7}$).

 We interpret these higher ratios as being due to shocks that are driven by the expansion of the bubbles.  Note that these enhancements are underrepresented in the \sii~and \oi~panels because the same shocks also enhance the \nii~emission, which is mixed in with \ha~here (see Appendix~\ref{wifesappendix} and Figure~\ref{wifes_lineratios} for pure line ratios at lower spatial resolutions).  As suggested by the \oiii~map itself, the \oiii/\hb~ratio map shows a sharp ridgeline in the eastern arm, likely pointing to a different ionization mechanism, which we explore below in Section~\ref{oiiiarm}.

We note that the \molhy/\pab~ratios are consistently high, showing a large peak encompassing both nuclei \citep[as identified by][providing a strong argument for \molhy~excited by merger-induced shocks]{vanderWerf93}, and enhancements along the filaments, similar to the optical line ratios.  For reference, \molhy/\brg~ratios in shock-excited gas tend to be $>>$1 \citep[e.g.][]{Puxley90}; using Case B recombination converts the threshold to \molhy/\pab~$>>$0.17 \citep{Hummer87}, showing that shocks dominate the excitation of \molhy~across the entire face of the galaxy.  The widespread nature of shock-excited \molhy~(elevated \molhy/\pab) is in contrast to the signatures of some outflows that produce shock-excited \molhy~in distinct cones (e.g. in NGC~4945, \citealt{Moorwood96,Marconi00}; in M82, \citealt{Veilleux09,Beirao15}; in IRAS F08572+3915, \citealt{Rupke13_08572}; in IRAS F17207-0014, \citealt{Medling15a}; in IIIZw035, \citealt{U19}, although even among those cones, the structure varies).  We examine the phase structure in the filaments of NGC~6240 in Section~\ref{bubbles} and explore the dust structure and its relation to the molecular gas in Section~\ref{dust}.

\subsubsection{Emission Line Diagnostics of Key Regions}
\begin{figure}
\centering
\includegraphics[scale=1.1,trim=2.6cm 1cm 1.5cm 1cm]{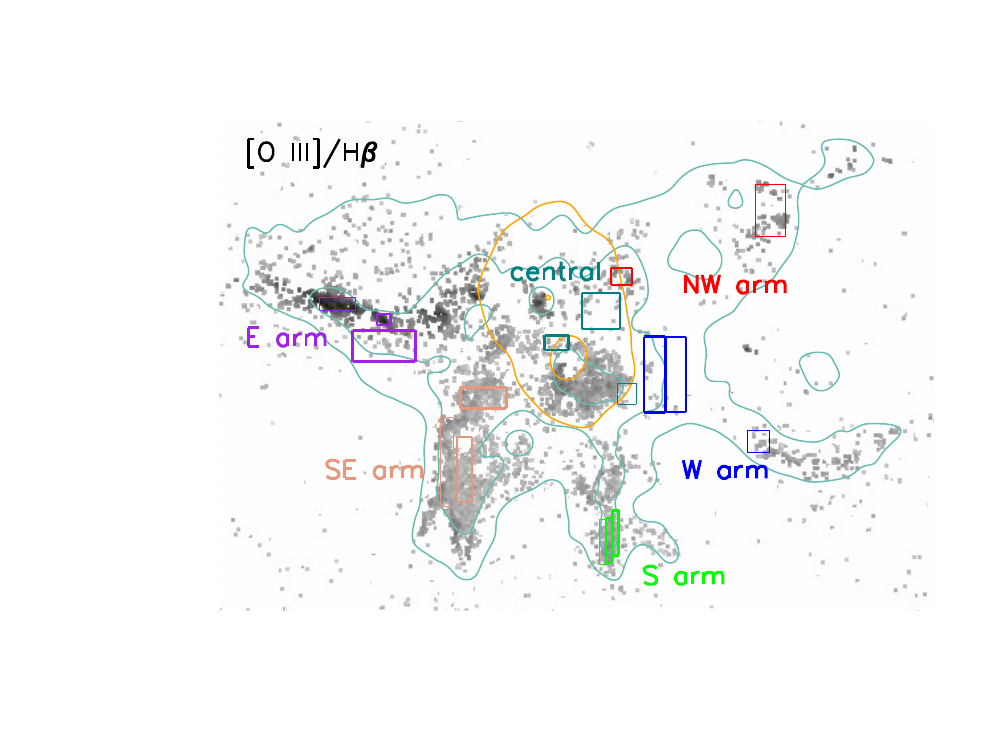}
\includegraphics[scale=1.1,trim=2.6cm 1cm 1.5cm 1cm]{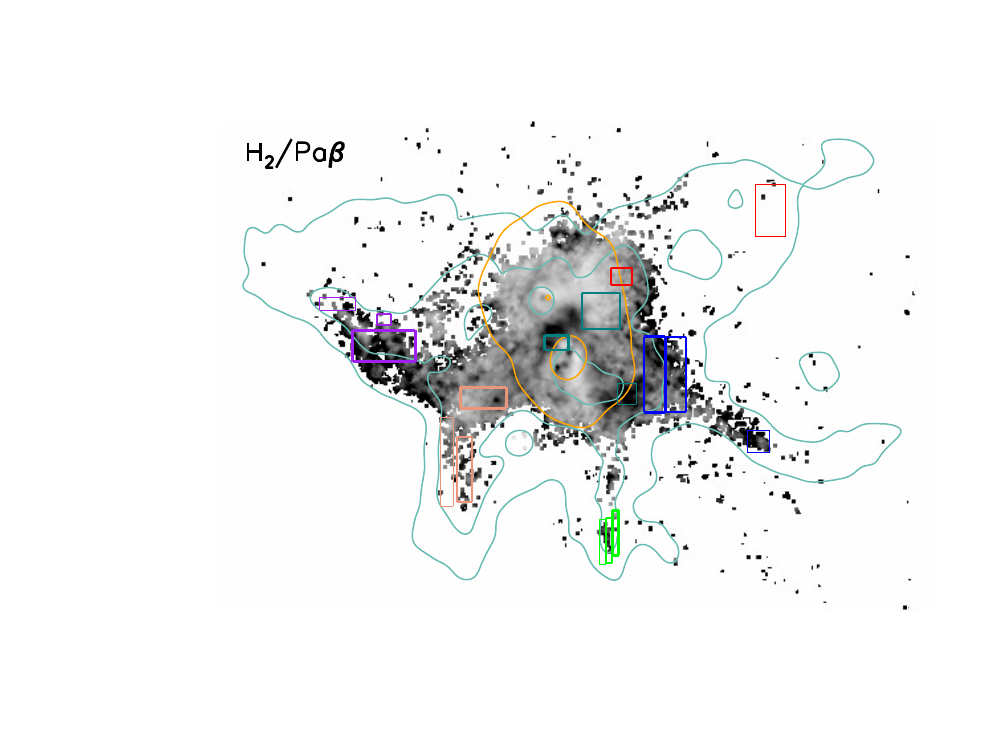}
\caption{Maps of log(\oiii/\hb) (top) and \molhy 1-0 S(1)/\pab~(bottom) from Figure~\ref{ratiomaps} with selected regions of interest boxed for comparison.  Line ratios within these regions are examined in Figure~\ref{BPTdiagnostics}.  The colors of the boxes here correspond to general areas of the galaxy, and match the colors in the diagnostic plots; within a color, the thinnest boxes correspond to the smallest symbols.
Each panel also has the \ha+\nii~contours (teal) and stellar continuum~contours (orange) from Figure~\ref{linemaps} overlaid for reference.  
}  
\label{regionlocator}
\end{figure}

To facilitate comparisons, we also examine line ratios in specific regions of interest.  Figure~\ref{regionlocator} shows our selected regions overlaid on the log(\oiii/\hb)~and \molhy~1-0S(1)/\pab~panels from Figure~\ref{ratiomaps} for reference.  For each region, we mask out spaxels below a 3$\sigma$ detection limit and then compute the resistant mean using a further 3$\sigma$ threshold.   The narrowband \nii+\ha~image was corrected based on the lower spatial resolution \nii/\ha~ratio from our WiFeS data to obtain a (smoothed) pure \ha~image that we used to measure these line ratios.

\begin{figure*}
\centering
\includegraphics[scale=1.2,trim=2.2cm 0.25cm 1.5cm 0.5cm]{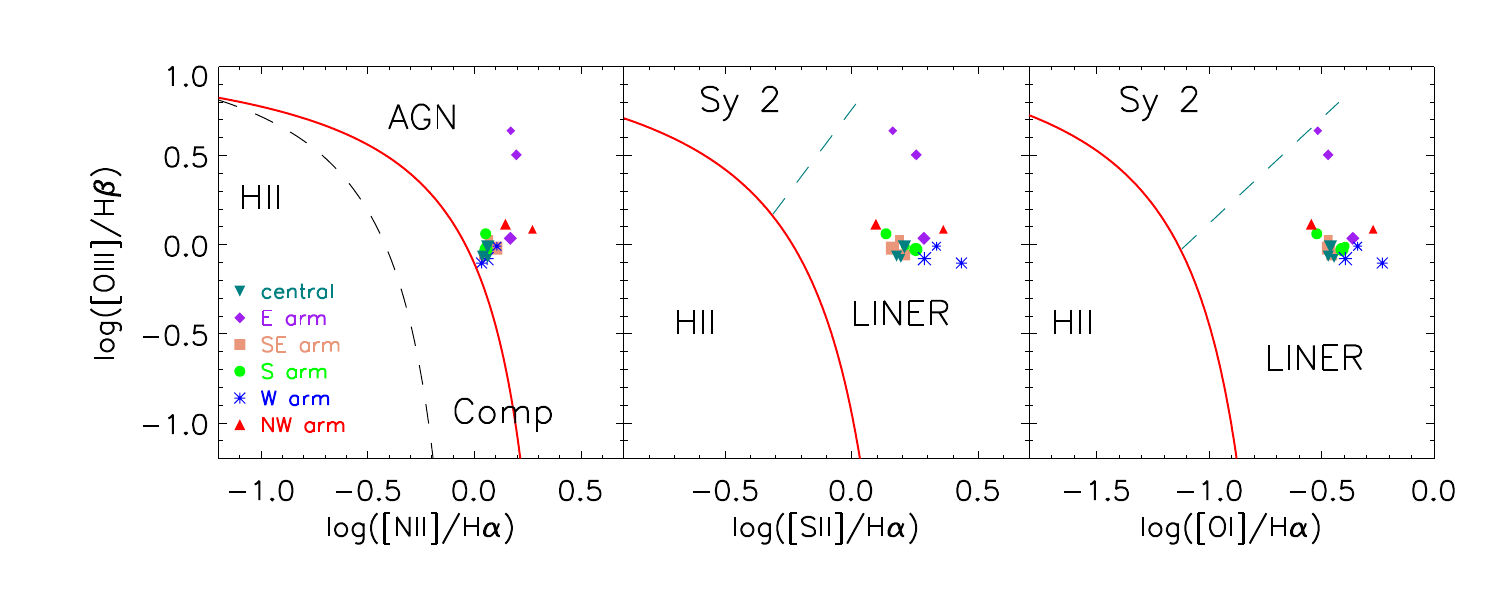}
\caption{Diagnostic diagrams for regions described in Figure~\ref{regionlocator}, following \citet{BPT, Veilleux87}.  Panels show (left to right) log(\oiii/\hb) vs. log(\nii/\ha), vs. log(\sii/\ha), and vs. log(\oi/\ha).  We include the classification lines for different photoionization mechanisms from \citet{Kewley06}.  Symbols are color-coded based on location in the galaxy: three central regions (teal downward-facing triangles), three from the east arm (purple diamonds), three from the southeast arm (peach squares), three from the south arm (green circles), three from the west arm (blue asterisks), and two from the northwest arm (red upward-facing triangles).  Each symbol is also size-coded to distinguish within a category: smallest symbols are furthest from the nucleus and have the thinnest lines in Figure~\ref{regionlocator}.  All regions are consistent with shocks/LINER-like ionization mechanism except the two regions along the ridge of strong \oiii, which are shifted towards AGN-like photoionization.
}  
\label{BPTdiagnostics}
\end{figure*}

As expected from Figure~\ref{ratiomaps}, the \oiii~ridgeline in the East arm (smaller two purple diamonds in Figure~\ref{BPTdiagnostics}) shows \oiii/\hb~ratios distinctly higher than all other points, leading them to stand out in most diagnostic plots.  The enhanced \oiii/\hb~in these two regions is almost but not quite sufficient to unambiguously place them in the AGN category.  We note that shock-ionization and AGN-photoionization models overlap in this line ratio space and further information is required to confirm the likely ionization mechanisms (which we explore in Section~\ref{oiiiarm}).  The remaining regions are relatively tightly clustered in the LINER or shock-like ionization mechanism regime.  It's interesting to note that the three central regions are among those that show the least AGN-like optical line ratios, demonstrating that the AGN are obscured from most of the surrounding ISM.

The selected regions also show a large spread (4x) in \molhy/\pab~ratio.  In \oiii/\hb, a single arm was distinct from the other arms; with \molhy/\pab, each filament (color/symbol) also spans a range.  We note that the outer regions of several arms show higher \molhy/\pab~ratios than other regions along the same arms.  An enhanced line ratio may be indicating stronger or more dominant shocks or simply less star formation (lower \pab~emission) along our line of sight further from the nucleus \citep[analogous to the increased optical line ratios in the outskirts of LIRGs;][]{Rich11}.  However, even the lowest ratios seen in our map (\molhy/\pab$\sim$2) are about an order of magnitude above the typical ratios expected from UV-pumping associated with young star formation \citep{Puxley90, Doyon94, Larkin98, Davies03}.

We see no sign of correlation between \molhy/\pab~and each of \nii/\ha~or \oiii/\hb; a weak correlation may exist between \molhy/\pab~and each of \oi/\ha~and \sii/\ha, but it is not statistically significant (both have Pearson's correlation coefficients $\sim$0.4).  Such a correlation would not be unexpected because all three line ratios are shock tracers.  

We also consider the effects of extinction on these shock diagnostics.  By using ratios of lines close in wavelength, we expect that extinction itself does not heavily affect our direct measurements.  However, the near-infrared light may be coming from physical regions more enshrouded by dust than the optical counterpart.  If most shocked filaments were buried in dust, we'd see a stronger shock signal in near-IR tracers than in optical tracers.  In NGC~6240, we see evidence of shocked line ratios in both optical and near-IR tracers in all regions probed in Figure~\ref{regionlocator}.  We can't rule out the possibility that some shocked gas is dust-enshrouded, but the presence of shocked optical line ratios tells us that some filaments aren't completely obscured.  Indeed, our Balmer decrement map (Figure~\ref{ratiomaps} shows the most extreme extinction around the two nuclei (blue, \ha/\hb$\sim$20), and most filaments show lower Balmer decrements than the regions between them (e.g. E and SE arms are pale red with values $<$10, between them is bright red with values of 10-15).


\section{Shock Structure in Bubbles}
\label{bubbles}

\begin{figure}
\centering
\includegraphics[scale=0.9,trim=0.2cm 1cm 0cm 0cm,clip]{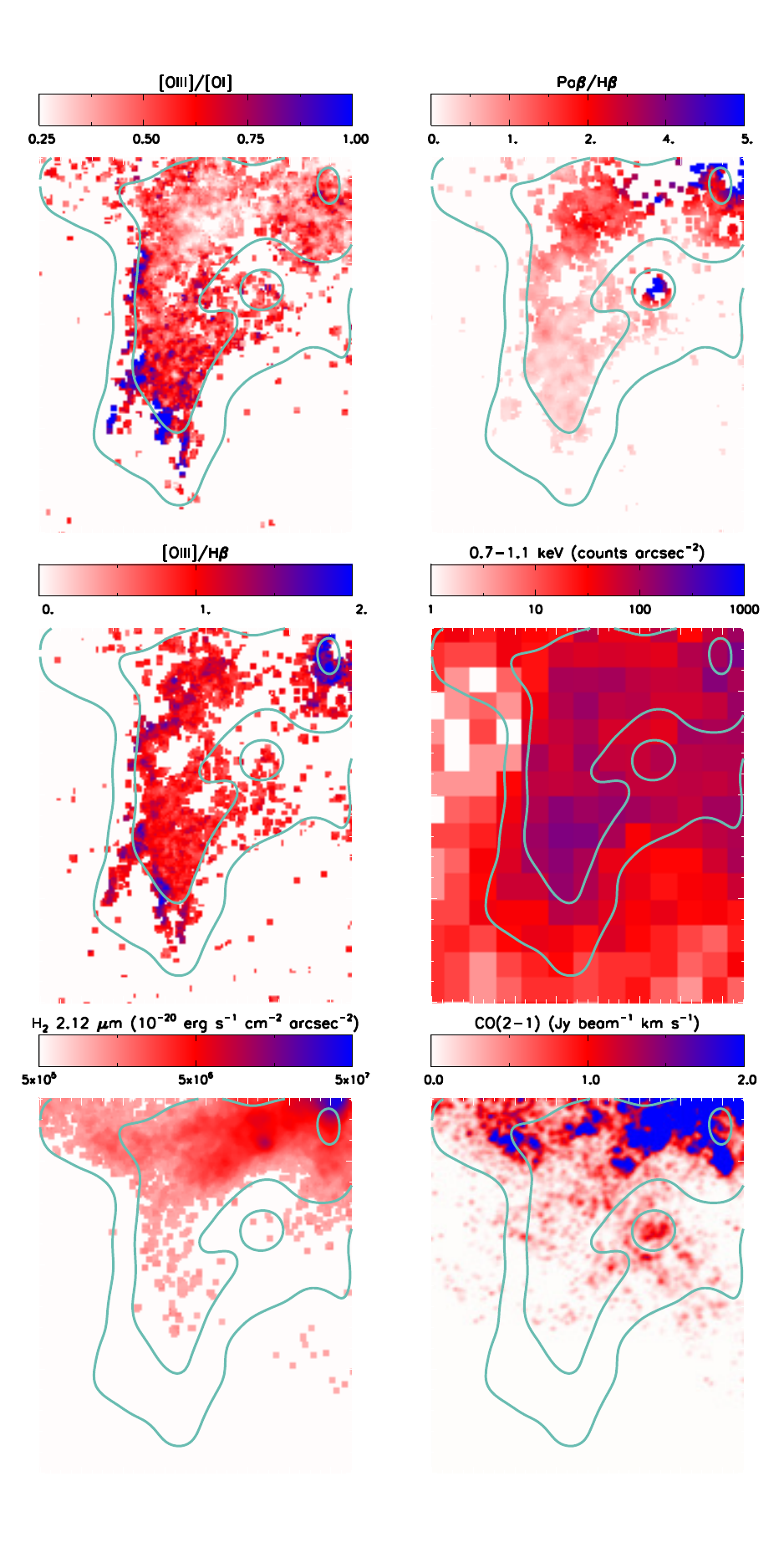}
\caption{Zoomed-in region of the southeast arm showing maps of \oiii/\oi~(top left), the \pab/\hb~decrement (top right), \oiii/\hb~(center left), soft X-ray (center right), \molhy~2.12$\mu$m (bottom left), and CO(2-1) (bottom right).  Teal contours show the \ha+\nii~flux for comparison between panels.  The \oiii/\oi~ratio is enhanced along the outer edge of the bubble (blue regions).  The \pab/\hb~ratio does not show a significant increase along the shell, so we are confident that the variation in \oiii/\oi~ratio is not due to variable dust extinction.  This enhanced ionization is also shown in \oiii/\hb and in the soft X-rays, confirming that the strongest shocks are present along the edge of the bubble.  \molhy~and CO emission are both present only in the interior of the bubble, and show no evidence for molecular gas reforming near the shock front.
}  
\label{SEbubble}
\end{figure}

As a bubble of hot gas expands, it can shock the surrounding interstellar medium and create increased emission line ratios along the edges \citep[e.g.][]{Weaver77}, exactly as we see here.  As an example, we zoom in on the southeast arm and examine the line maps in more detail (Figure~\ref{SEbubble}).  
In this region, we propose that gas is outflowing from the nuclei (beyond the top right corner of each panel) in a south or south-east direction.  The shock front would have expanded from the top right, moving across the panels until reaching its current location at the edge of the bubble.
The \oiii/\oi~ratio probes the ionization structure directly: we see a significant enhancement of \oiii~relative to \oi~along the outer edge of the shell, indicating the current location of the shock front.  That is, the high ionization state of the gas here indicates that it is hotter and therefore more recently-shocked than gas to the west, which has had longer to cool.

Because \oi~at 6300\AA~is significantly redder than the \oiii~5007\AA~line, this ratio could be subject to variable dust extinction.  However, if less dust extinction were causing the factor of 2-4 increase in \oiii/\oi~ratio we see along the outer edge of the shell, the \pab/\hb~map should show a corresponding decrease by at least 30\% in that region, which it does not.
\footnote{We do note that the decreased \oiii/\oi~near the top right of the panel is likely due to the increased dust extinction, traced by a higher \pab/\hb~ratio there.}  Indeed, the \oiii/\hb~ratio is less subject to extinction and shows a similar increase along that shell.  We note that this shell feature is unlikely to be an artifact of a variable point spread function across the relevant emission lines because the stars and star clusters used to register the different HST filters have statistically indistinguishable distributions of full-width at half-max.
At lower spatial resolution, the soft X-ray emission is spatially consistent with the edge of the bubble, although perhaps more aligned with the \ha+\nii~contours than with the higher ionization \oiii/\hb~emission.

The small \ha~peak in the middle of the bubble highlighted in Figure~\ref{SEbubble} is likely a young star-cluster located in front of the bubble.  It shows higher \pab/\hb~ratios ($\sim$5) than the surrounding areas, suggesting it is still dust-enshrouded, and is colocated with a clump of cold molecular gas emitting CO(2-1).  As with other star clusters, this clump shows no detected warm \molhy; it may even be obscuring some warm \molhy~emitted by the bubble behind it.

The presence of molecular gas in galactic winds \citep[e.g. ][]{Sturm11, Feruglio13a,Feruglio13b,Spoon13, Veilleux13,Cicone14, GonzalezAlfonso17, Gowardhan18} presents a theoretical challenge.  We discuss three possible scenarios here:

\begin{enumerate}
\item \textbf{Molecular Gas Is Entrained:} Some high-resolution simulations of molecular clouds entrained in hot outflows suggest that the clouds might be quickly shredded and evaporate into the hot phase of the wind \citep{Scannapieco15, Bruggen16, Schneider17}.  If molecular clouds are instead swept up and carried away by outflowing gas, we expect the molecular gas to be evenly distributed across the face of an outflow. Depending on the density and size of the clouds, they would eventually evaporate into the wind and 
become less frequent farther from the launch of the outflow \citep{BandaBarragan21}.  In this scenario, gas on the edges of each cloud may still be shocked even as the interior is shielded.

\item \textbf{Molecular Gas Is Dissociated and Reforms:}
An alternative theory proposes that the shock front sweeps through and dissociates all molecular gas, after which the initially-hot winds cool adiabatically and/or radiatively as they expand until new molecular clouds form from remaining density fluctuations or behind the shock front \citep{Thompson16, Schneider18} beyond a certain cooling radius.  If the outflowing molecular gas is (re-)formed, we might expect to see a higher incidence of molecular gas along the edges of a bubble.  However, the location of this putative re-formed molecular hydrogen depends on the cooling time/length, and could lag behind the shock front so far that we would not see its evidence in this particular bubble.
\citet{Ohyama03} favored a similar scenario to describe the shocked \molhy~emission between the two nuclei of NGC~6240, suggesting that an outflow from the south nucleus shocks the already-existing molecular gas bridge between the nuclei, crushing molecular clouds along the shock front.  We note that this scenario suggests that the lack of molecular gas in the middle of the bubbles is purely due to the shock having dissociated it; consequently, you'd expect molecular gas \textit{outside} the shock front, on the left edge of these panels.  

\item \textbf{Molecular Gas is Avoided:}
The simulations of \citet{Wagner13} show a third possibility: that the hot outflow mostly avoids the dense clumpy molecular clouds, forging pathways around them.  In this scenario, the distribution of the molecular gas is mostly unchanged from a pre-outflow scenario: that is, likely higher density closer to the nuclei (as we see).  The fact that the shock front is visible outside of the region containing molecular gas demonstrates that the hot wind was able to find pathways between the molecular gas and continues to push aside the remaining ISM.  As with the first scenario, gas on the edges of each cloud may still be shocked, producing signatures such as warm \molhy~2.12 $\mu$m emission and high \molhy~/\pab~ratios.
\end{enumerate}

In Figure~\ref{SEbubble}, the CO(2-1) and \molhy~emission fade quickly with distance from the nuclei and is here detected solely interior to the shocked shell of the bubble traced by enhanced \oiii/\hb~and \oiii/\oi~ratios).  The lack of CO(2-1) or \molhy~emission beyond the shock front traced by high \oiii/\oi~(i.e. on the left edge of the panel) suggests against the second scenario.  We argue then that the molecular gas seen inside the bubble is not gas that has been dissociated-and-then-reformed.  Still, if the cooling time/length were as short as in some other merging systems \citep[$\sim$10$^5$ yr, $\sim$few hundred pc in Stephan's Quintet;][]{Guillard09}, it ought to be possible here if gas densities are high enough.  We note that 
some other filaments of \molhy~(e.g. the west arm, between the nuclei, and below the east arm) show increased \molhy/\pab~ratios.  It is possible that conditions are different enough there to enable molecules to reform below or behind a shock front, or that something else is going on entirely, like the collision of two spiral arms along our line of sight.  Further, it is possible that the shock front was responsible for dissociating molecular gas previously, when the shock front had not yet expanded beyond the molecular-gas-rich nuclear region, but now is no longer overtaking substantial amounts of molecular gas.

Our data cannot distinguish between the first and third scenarios.  Both scenarios predict that the molecular gas decreases in density along the length of the outflow, can be present inside the bubble, would exhibit some warm \molhy~emission, and need not be present outside the shock front. 
Future analysis comparing the kinematics of CO(2-1), warm \molhy, and ionized gas emission lines are needed to determine how/if the molecular gas is moving along with the outflow.


\section{The Spatial Structure of Molecular Gas and Dust}
\label{dust}

Most of the molecular gas traced by CO is in a ribbon or bridge between the two nuclei \citep{Tacconi99}, although significant CO(2-1) emission is seen extending $\sim$6.5\arcsec ($>$3 kpc) from the nuclei in several directions.  Our narrowband \molhy~emission also traces this bridge, similar filamentary structure within the central kiloparsec, and the extended arm to the southeast.  We detect no warm \molhy~component in the north or south-southwest CO extensions.  Interestingly, we do see warm \molhy~in the southwest arm and further along the east arm that appears to have only a much weaker CO counterpart.

\begin{figure}
\centering
\includegraphics[scale=0.34,trim=0cm 4.5cm 0cm 4.5cm,clip]{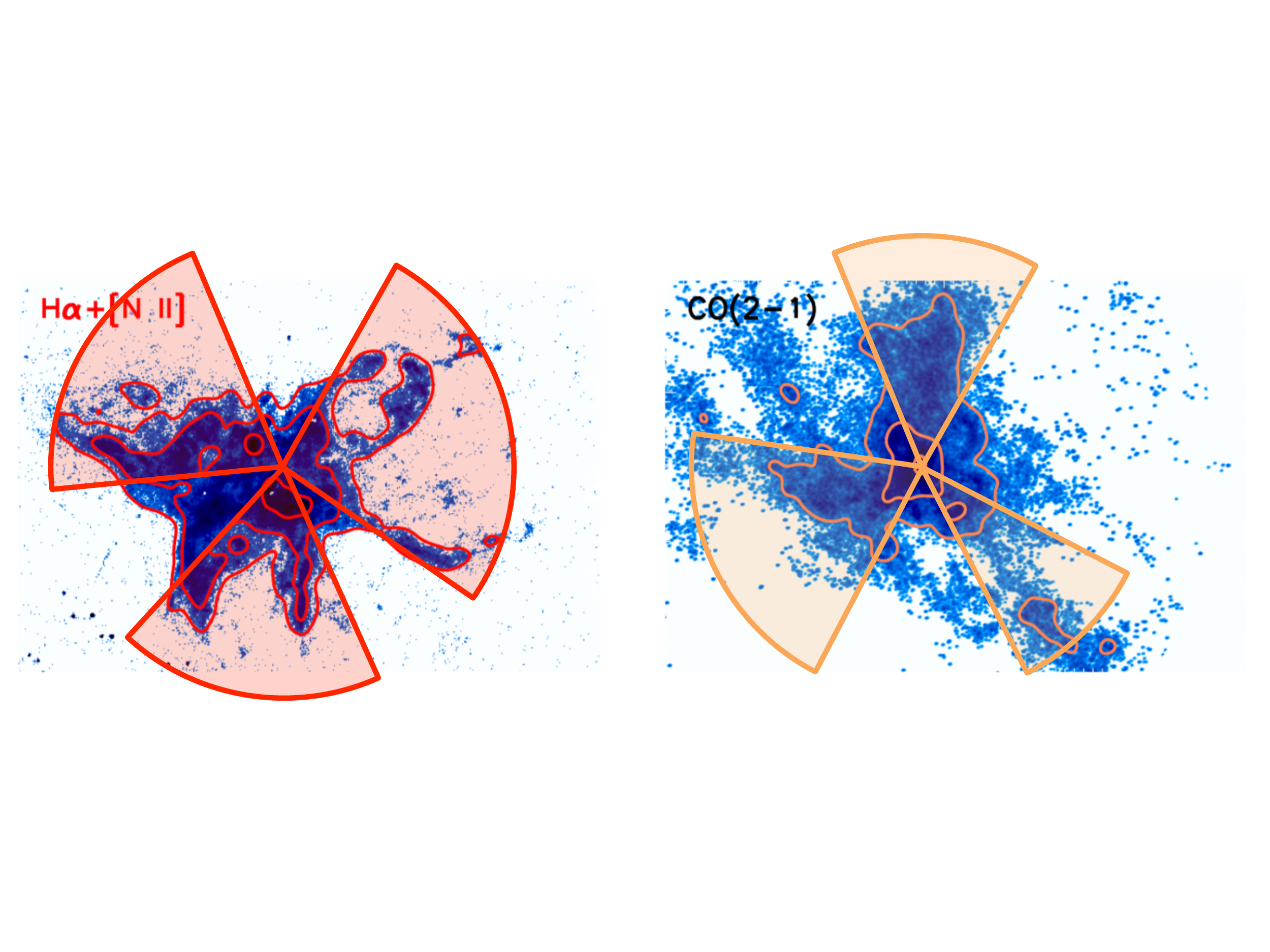}
\caption{\ha~(left) and CO(2-1) (right) emission line maps from Figure~\ref{linemaps}, reproduced with their flux contours in red and orange, respectively.  In each panel, we overlay wedges extending from the nucleus along regions of increased ionized or molecular gas, to highlight that the two phases of gas are qualitatively anticorrelated.  Such an anticorrelation may indicate that the hot ionized gas is flowing outwards along the path of least resistance, in between clumps of denser molecular gas.
}  
\label{anticorr}
\end{figure}

Feedback in a galaxy with a multiphase ISM involves the hot gas expanding outwards symmetrically until it reaches something to block it.  Dense clumps of molecular gas are harder to accelerate, so the gas may expand around them, following a path of least resistance \citep[e.g.][]{Wagner13}, just as in the third scenario of Section~\ref{bubbles}.  The molecular gas structure here is reminiscent of that scenario: extended / clumps of CO(2-1) to the north, southeast, and southwest all align well with spaces \textit{between} the bubbles traced by ionized gas filaments (at least in projection; of course the three-dimensional structure may be more complex).  We highlight the spatial anticorrelation between \ha~and CO(2-1) emission in Figure~\ref{anticorr}.

Our Balmer decrement map, shown in Figure~\ref{ratiomaps}, demonstrates that the bulk of the dust extinction also appears around or between the nuclei, with moderate increases towards the southeast and southwest arms, spatially coincident with the CO(2-1) emission and particularly its north extension.  We note that the actual extinction could be higher in places because we are only tracing the regions where \ha~and \hb~are optically thin.
The presence of dust may be necessary to shield the \molhy~from dissociation or we may simply be seeing evidence of a constant gas-to-dust-mass ratio, wherein denser regions of gas happen to also host more dust mass.  

\begin{figure*}
\centering
\includegraphics[scale=0.95,trim=3.5cm 0.5cm 1.5cm 0cm]{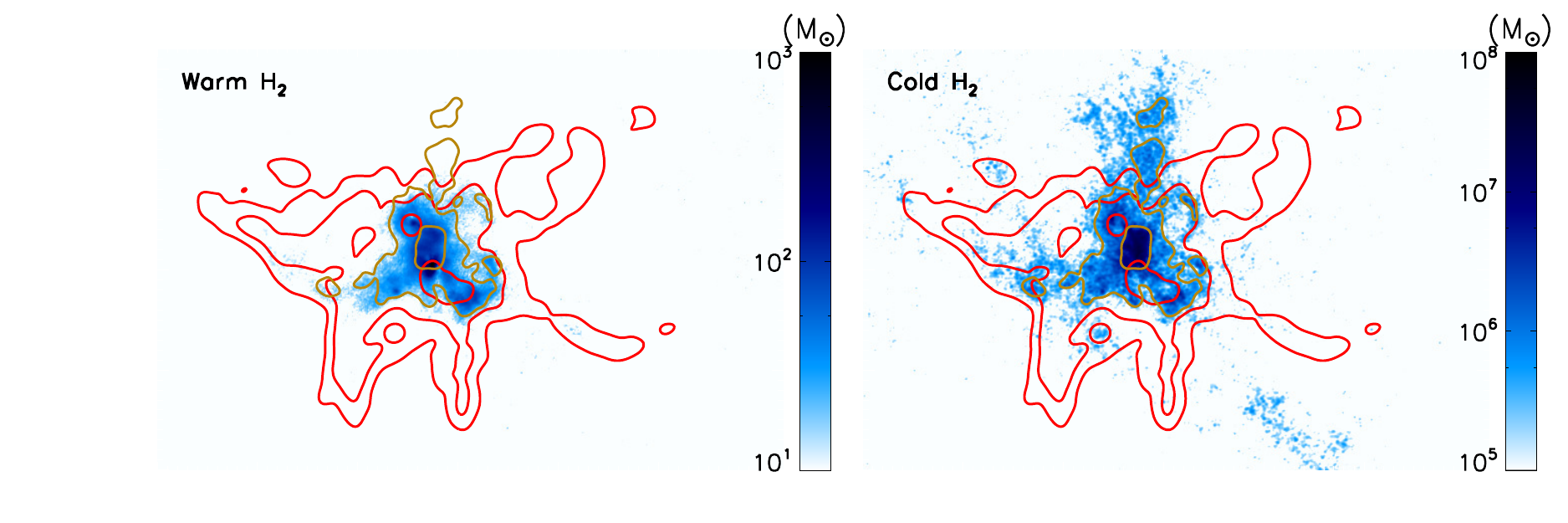}\\
\includegraphics[scale=0.95,trim=3.5cm 0.5cm 1.5cm 0cm]{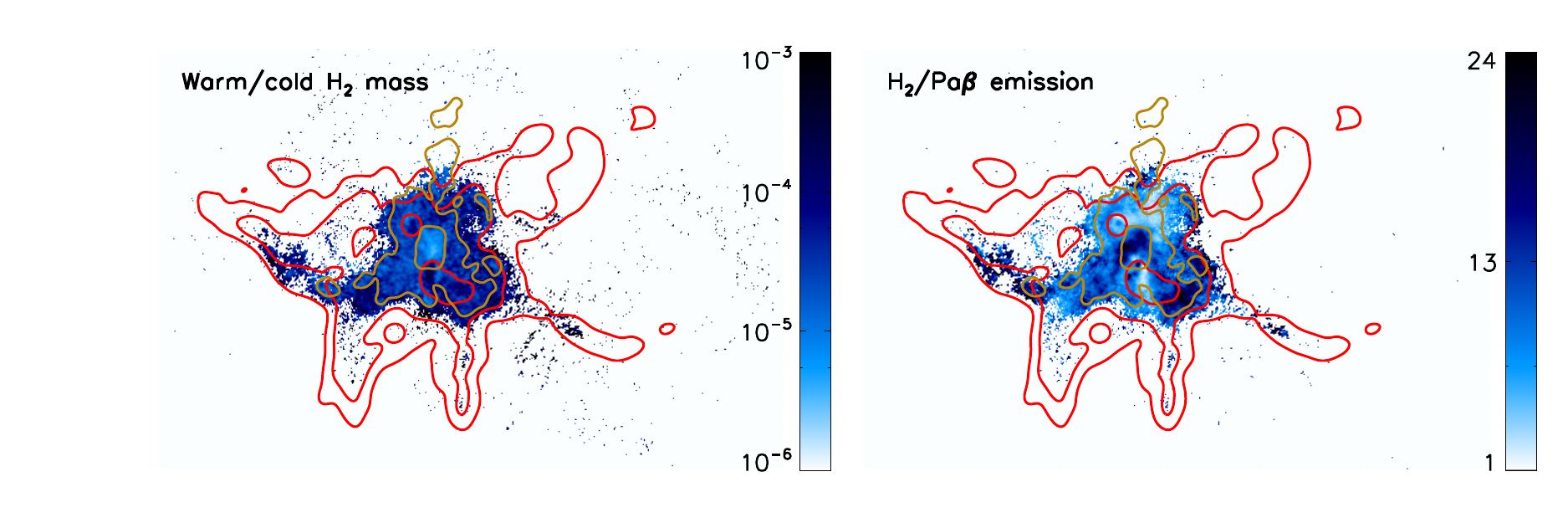}
\caption{Top: Mass maps (in M$_{\sun}$ \text{pixel}$^{-1}$) of warm (T$\sim$2000K) \molhy~calculated from \molhy~2.12 $\mu$m emission and of cold \molhy~calculated from CO(2-1) emission.  Bottom: \molhy/CO(2-1) and \molhy/\pab~ratio maps with \ha+\nii~(red) and CO(2-1) (gold) contours for comparison.  
The mass fraction of warm molecular gas increases to $\sim10^{-4}$ in the filaments, 1-2 orders of magnitude larger than in the shocked bridge between the nuclei.
}
\label{molgas}
\end{figure*}

We calculate the mass present in warm \molhy~from the \molhy~2.12 $\mu$m emission line following \citet{Veilleux09}.  Assuming the \molhy~molecules are thermalized at T=2000K, M$_{H_{2}}$ = 0.00133 [L$_{S(1)}$/L$_{\sun}$] M$_{\sun}$.  For comparison, we also calculate the mass of cold \molhy~traced by our CO map using the mean $\alpha_{\text{CO}}$ = 2.5 and $r_{21}$ = 1.17 measured across NGC6240 by \citet{Cicone18}.  The resulting maps (Figure~\ref{molgas}) show that, in the large ribbon of molecular gas between the nuclei, only a small fraction ($\sim$3$\times$10$^{-6}$) is excited by shocks or ultraviolet photons; however, in the filaments, the fraction is roughly 100 times higher.  For comparison, the outflow in M82 shows an integrated warm/cold (from \molhy~2.12 $\mu$m relative to CO(2-1)) gas mass ratio of $3\times10^{-5}$ \citep{Walter02,Veilleux09}, an intermediate value in the range we see here.  We also estimate the rough fraction of warm to cool gas expected in a pure star-forming region as follows: the Orion Nebula has \molhy/\brg~ratio of $10^{-1.2}$; we convert the \molhy~to mass as above, and convert the \brg~luminosity to a star formation rate and the star formation rate (in unit surface area) to a gas mass (in the same unit surface area) following relations in \citet{Kennicutt98}.  We thus estimate the mass fraction traced by \molhy~2.12 $\mu$m to that traced by CO(2-1) as roughly 10$^{-6}$ in a pure star-forming system.  As an independent constraint, we extrapolate the warm \molhy~fractions seen in the top left panel of Figure~7 in \citet{Roussel07} for SINGS galaxies; when defining `warm' \molhy~as the T$>$2000K gas traced by our 2.12 $\mu$m transition, a pure star-forming region should have a warm-to-cold \molhy~fraction of $\sim$ 2.5$\times10^{-6}$.  Both estimates of warm-to-cold \molhy~fractions in normal star-forming regions are below most of the map in the bottom left panel of Figure~\ref{molgas}.  

The sometimes substantial extinction can affect these ratio maps, as follows.  In regions with high extinction (i.e. around the nuclei and extending north), the \molhy~2.12$\mu$m emission will be more attenuated than the longer-wavelength CO(2-1), artificially depressing the warm to cold gas mass ratio maps.  Further, the \molhy~2.12$\mu$m emission will be less attenuated than the \pab~emission at shorter wavelengths, potentially inflating the subsequent line ratio map.  Our Balmer decrement map from Figure~\ref{ratiomaps} suggests A$_{\text{V}}$ ranges from $\sim$1 to $>$100 with a strong spatial dependence.  Because we only have a measure of high extinction at low spatial resolution, we do not apply a direct correction to these maps, and note that the warm to cold gas ratio measured here is likely an underestimate.

We also note that the fraction of warm-to-cold molecular gas depends heavily on the definition of `warm' and thus on the observed transitions of \molhy~emission; mid-infrared lines will trace cooler gas (100-500K) that represent a larger mass of shocked gas \citep[see e.g. M82 in][]{Beirao15}.  Future analyses mapping both near- and mid-IR transitions of \molhy~at comparable spatial resolutions will be possible with JWST in the coming years and (along with ALMA dense gas mapping) will reveal the temperature and shock structure of the molecular gas.  

The strong 2.12$\mu$m emission along the ribbon between the nuclei has been attributed to intense shocks due to the ongoing collision of the two nuclei \citep[e.g.][]{Tecza00}, and the region's high CO line-to-continuum ratio may require strong shocks \citep{Meijerink13}.
Interestingly, we see no increase in the fraction of molecular gas that is shock-excited in this region between the nuclei.  \citet{Ohyama03} also found an increase in 2.12$\mu$m production efficiency moving away from the nuclei (primarily towards the southwest).
The spatial variation in fraction of excited \molhy~is qualitatively similar to the \molhy/\pab~map also shown in Figure~\ref{molgas}, where the ratio is higher further from the nucleus, with the notable exception of a very high \molhy/\pab~ratio in the bridge between the nuclei.  The molecular gas in the bridge, therefore, is not necessarily too warm to form stars, but star formation is nonetheless suppressed, perhaps by turbulence.  A full analysis of the CO kinematics at these spatial resolutions will be presented in a forthcoming paper.  The increasing fraction of shock-heated molecular gas further from the nucleus has been seen frequently in optical shock studies, and is usually interpreted as the star formation decreasing rather than shocked gas increasing \citep[e.g.][]{Rich11}.  Our comparison of warm to cold molecular gas may provide a more direct measure of the spatially varying state of the gas, and confirms that shocks are more dominant in the filaments.

\section{Collimated \oiii~Emission}
\label{oiiiarm}

\begin{figure}
\centering
\includegraphics[scale=1.1,trim=3.5cm 1.5cm 1.5cm 0.25cm]{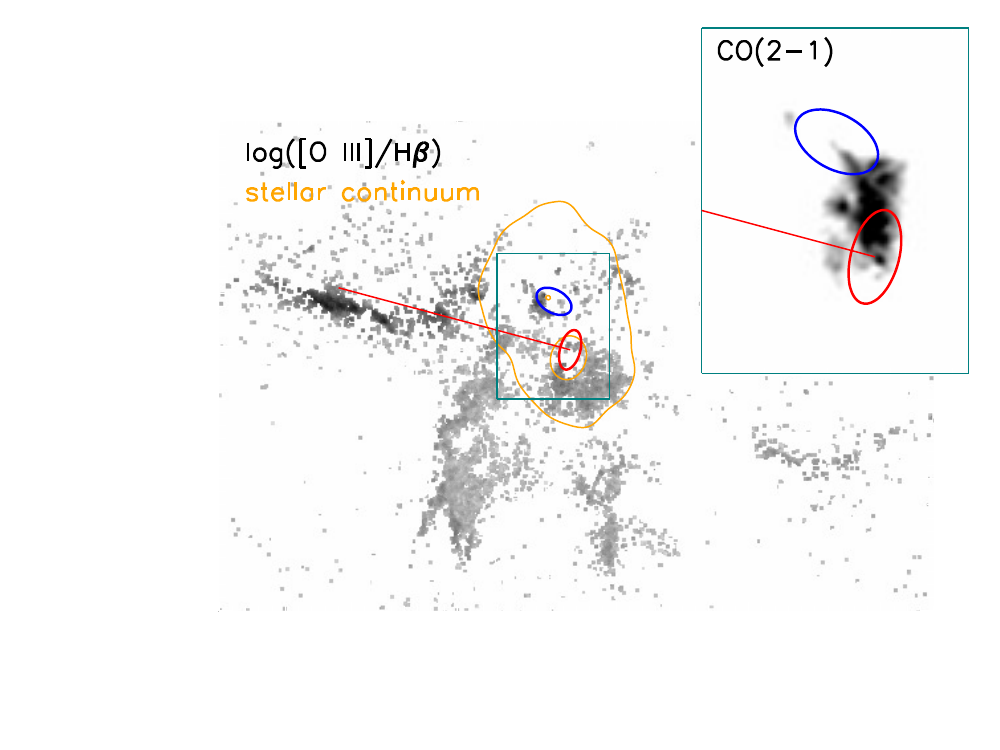}
\caption{Schematics of north (blue) and south (red) nuclear disks modeled in \citet{nucleardisks} overlaid on log(\oiii/\hb) map.  Axis ratios and position angles are as fitted, but sizes are enlarged for clarity.  Red line shows projected minor axis of southern disk, closely aligned with the high \oiii/\hb~ridgeline.  Stellar continuum contours are shown in orange.  The inset shows a zoom-in of the teal box with grayscale of CO(2-1) emission and schematics overlaid, highlighting the possible ring structure along the line between the south nucleus and the high \oiii/\hb~ridgeline.}  
\label{diskcartoon}
\end{figure}

The two regions from Figures~\ref{regionlocator} and \ref{BPTdiagnostics} that fall along the collimated \oiii~emission show high log(\oiii/\hb) ratios ($>$0.5) but are unremarkable in all other emission line ratios we considered.  These high log(\oiii/\hb) ratios suggest that gas along this region is ionized by a different mechanism than the other filaments present in this galaxy.  \citet{MuellerSanchez18} identified this \oiii~emission as the edge of an ionization cone.  The west arm is aligned with our ridgeline on the opposite side of the nuclei; it may be a counterpart to the ionization structure, although the \oiii/\hb~and other line ratios are more consistent with the other filaments than with the east arm.  The southeastern extension of molecular gas loops north, becoming more shocked as it approaches the ridgeline of high log(\oiii/\hb), although it does not extend far enough to overlap.
We see two possible scenarios explaining the ridgeline: either this region shows a collimated jet from one of the AGN shock-heating the gas or a direct line-of-sight path has enabled this gas to be photoionized from the AGN's radiation field.  

In either case, the energy source is likely one or both of the two AGN.  The length of the \oiii~ridge points directly to the south nucleus.  Indeed, the projected line-of-sight aligns with the axis of rotation of the nuclear disk surrounding the southern AGN \citep{Medling11, nucleardisks}, with only a small offset (Figure~\ref{diskcartoon}).  A small ring-like structure in the CO(2-1) is visible just east of the south nucleus, along its minor axis, along the projected path to the east arm.
Although the high log(\oiii/\hb) ridgeline extends almost due east from the northern AGN, the position angle of that nuclear disk points S-SE, so an AGN-driven jet or beam of radiation would have to emerge closer to the plane of the disk.

\begin{figure*}
\centering
\hspace{.8cm}\includegraphics[scale=0.95,trim=3.5cm 0.5cm 1.5cm 0cm]{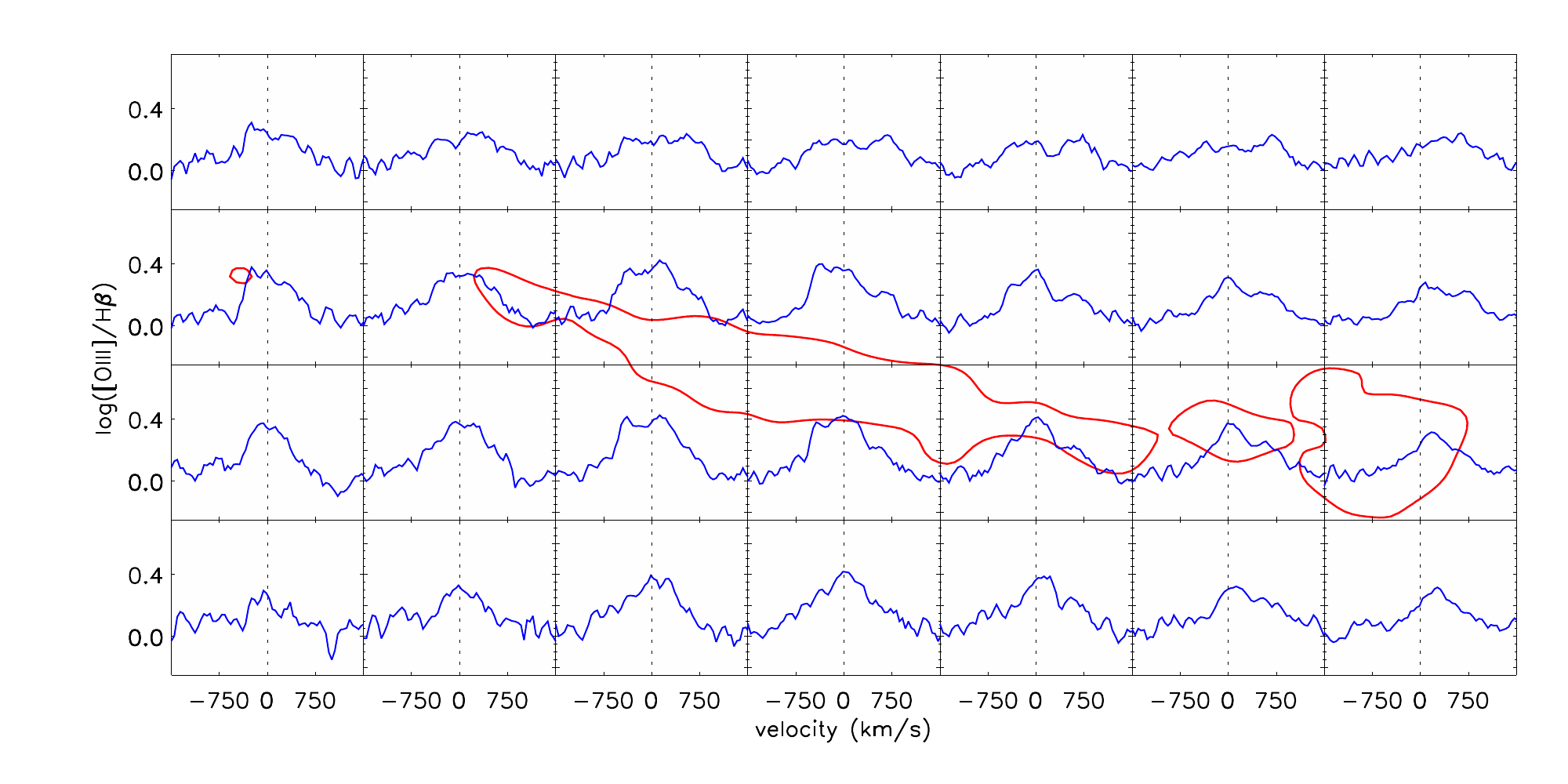}\\
\caption{
log(\oiii/\hb)~ratios as a function of velocity for the 7\arcsec$\times$4\arcsec region around the ridge taken from our WiFeS integral field spectroscopy, with each panel corresponding to an individual WiFeS spaxel.  The emission line ratios have been smoothed over three spectral channels to better show the velocity structure.
For comparison, contours from the HST-derived log(\oiii/\hb) map in Figure~\ref{ratiomaps} covering the same spatial region are shown in red.  Note that seeing will smear the signatures of the ridge out over multiple WiFeS spaxels.  Across the whole spatial region, the \oiii/\hb~ratios show similar kinematic structure: centered around systemic velocity; lines are broad (several hundred km s$^{-1}$ everywhere).  The high \oiii/\hb~gas does not show the velocity gradient or strong velocity offset that would be expected from an outflow (unless it is in the plane of the sky).
}  
\label{OIIIarmlineprofiles}
\end{figure*}

In Figure~\ref{OIIIarmlineprofiles}, we show the line profiles from our WiFeS integral field spectroscopy for a 7\arcsec$\times$4\arcsec region around the ridge.  The red contours in the background show the spatial region over which the high \oiii/\hb~ridge would contribute if beam smearing were not an issue.  In these data, however, the 1\farcs5 seeing has smeared the spectral signatures of the ridge out over most spaxels in the given region.  
These spectra show only minor differences between the kinematic structures of \oiii~and \hb.  The minor enhancement in \oiii~over \hb~in spaxels dominated by the ridgeline is centered at systemic velocity.  The lack of a strong velocity offset of the high \oiii/\hb~gas argues against the AGN-driven outflow interpretation, although if the outflow is mostly in the plane of the sky, its velocity offset would be hard to detect.  The broad nature of both \oiii~and \hb~are consistent with a highly turbulent outflow, but also could be due to beam smearing of a complex kinematic structure; the multi-peaked nature of the individual line profiles suggests the latter.

\begin{figure*}
\centering
\hspace{.8cm}\includegraphics[scale=0.95,trim=3.5cm 0.5cm 1.5cm 0cm]{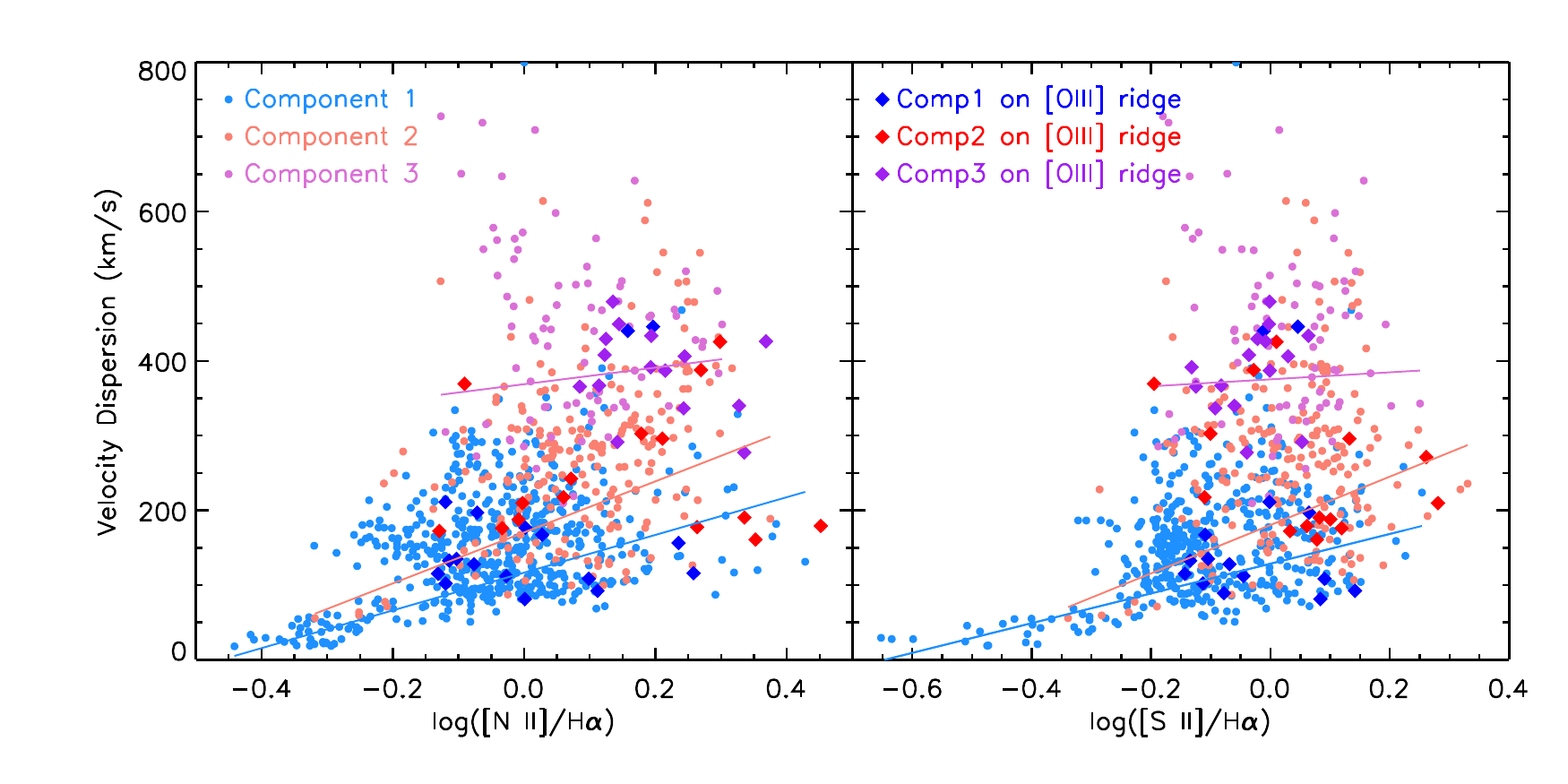}\\
\caption{Velocity dispersion as a function of \nii/\ha~(left) and \sii/\ha~(right) ratios for each of the three Gaussian components (sorted from narrow to broad) fit for all spaxels from the WiFeS data meeting a signal-to-noise ratio of 5 or higher in relevant emission lines.  Bold-colored larger diamonds represent spaxels appearing along the \oiii~ridgeline.  Overall, the 
two narrower components in each spaxel
show statistically significant positive trends for both line ratios, a key signature of shocks, while the broadest component does not.  The spaxels lying along the high \oiii~ridgeline show no positive trends in any component.
}  
\label{lineratiovdisp}
\end{figure*}

If the \oiii~arm traces out a collimated jet or other outflow signature, we would expect evidence of shocks across all tracers.  However, we see no evidence of enhanced \sii/\ha, \oi/\ha, \molhy/\pab, \fexxv, or soft X-rays along this ridgeline in our maps, and the two regions standing out so highly in \oiii/\hb~in Figure~\ref{BPTdiagnostics} are not elevated in other line ratios.  
Shocked gas often displays a positive correlation between line ratio and velocity dispersion \citep[e.g.][]{Rich11,DAgostino19,Kewley19}, which we see across the entirety of NGC~6240 in Figure~\ref{lineratiovdisp}.  However, the spaxels along the \oiii~ridgeline do not show such a correlation, further arguing that its ionization is not shock-dominated.  We also consider the possibility that some excess ionization is due to a fully-ionized shock precursor here that isn't present elsewhere in the galaxy.  \citet{Sutherland17} find that shock precursors transition to fully-ionized around shock velocities of $>$140 km s$^{-1}$, but we see no evidence that the \oiii~emission has a stronger velocity offset (Figure~\ref{OIIIarmlineprofiles}) or broader velocity dispersions (Figure~\ref{lineratiovdisp}) than the shocks present across the majority of the galaxy.  However, we note that velocity dispersions across the galaxy, if they are due to shocks, are easily capable of photoionizing precursor gas everywhere.  We can't rule out the possibility that the E arm ridgeline represents a unique location where we get the best view of said precursor gas radiating, although we also see no particular evidence to support it (i.e. higher gas densities).

If the putative jet were strong enough to drill through the ISM without stirring it up, we would expect to see evidence in the radio, but no radio emission is evident along the east arm \citep{Colbert94}, although a relic radio feature suggests a jet may once have pointed in this direction \citep{Yun99}.  On the other hand, if we were looking at beam of ionizing radiation from an AGN, we would expect the ionization parameter to decrease with radius unless the density also decreases.  We do not see a trend in \oiii/\hb~along the $\sim$300 pc of the ridgeline, although we also do not have a measurement for the density of hydrogen along the same spatial scales.  (We note that the molecular gas mass surface density from our CO(2-1) emission does not show a significant trend along the ridgeline.)

We summarize the evidence in support of each scenario in Table~\ref{tbl:oiiiarmevidence}.  Although not conclusive, we find the evidence more in favor of a beam of AGN photoionization causing the enhanced \oiii/\hb~in the East arm.

\begin{table*}
\caption{\oiii~Arm Evidence}\label{tbl:oiiiarmevidence}
\begin{tabular}{lc}
\hline
\hline
Evidence & Y/N \\
\hline
\hline
\textbf{Scenario I: Collimated AGN Jet or Outflow} &\\
Elevated \oiii/\hb & Y \\
Aligns with a nucleus & Y \\
Optical line ratios consistent with shocks & Y \\
Optical line ratios correlated with velocity dispersion & N \\
Elevated \molhy~2.12$\mu$m, \fexxv, soft X-ray emission & N \\
Radio emission tracing the jet/ridge & N \\
Enhanced optical line widths & N \\
\hline
\textbf{Scenario II: AGN Photoionization through a Keyhole} \\
Elevated \oiii/\hb & Y \\
Aligns with a nucleus & Y \\
Optical line ratios consistent with AGN & some \\
 ---- or consistent with AGN + shock mixing & Y \\
Optical line ratios uncorrelated with velocity dispersion & Y \\
Unremarkable optical line widths & Y \\
Decreasing \oiii/\hb~with AGN distance (if constant $n_e$) & N \\
\hline
\end{tabular}
\end{table*}

\section{Conclusions}
\label{conclusions} 

We present high spatial resolution emission line maps from the Hubble Space Telescope, ALMA, and the Keck Observatory, and optical integral field spectroscopy from the Wide Field Spectrograph of the Butterfly Nebula of NGC~6240.  Our dataset enables a detailed comparison of the multiphase energetics in and around the filaments and bubbles associated with starburst- and AGN-driven winds.  We find:
\begin{itemize}
\item The bubbles inflated by the superwind expand outwards along the paths of least resistance, avoiding regions with the highest densities of molecular gas.
\item Shocked molecular gas is seen along most but not all filaments and is generally interior to the outer shock fronts traced by optical ionization line ratios.  This morphology suggests that either molecular gas is entrained in the outflows and evaporates within a few kiloparsecs, or that it is not entrained at all.  We do not see evidence for molecular clouds being \textbf{dissociated} and then reforming behind shock fronts, although we do note that the phase structure varies between filaments.

\item We find a ratio of warm \molhy~(T$\sim$2000K, traced by the 1-0 S(1) transition at 2.12 $\mu$m) to cool \molhy~(traced by the CO(2-1) transition) of $\sim$5$\times10^{-6}$ by mass between and around the nuclei, with no obvious increase in shock-excited gas in the ribbon between the two nuclei.  However, towards the edges of the nuclear region and along the filaments, the fraction does increase to roughly $10^{-4}$.  Cloud-crushing in the bridge associated with the two nuclei merging is therefore not a significant cause of shock-excitation in NGC~6240 compared to outflows.
\item A narrow ridge of elevated \oiii/\hb~emission in the eastern arm aligns spatially with the minor axis of the south nucleus' disk.  This emission is likely either a tightly-collimated outflow from the AGN or a pencil-beam of radiation that found a path through the otherwise thickly-obscuring interstellar medium.  The highly ionized gas shows none of the kinematic signatures of an outflow, favoring the latter scenario.
\end{itemize}

\acknowledgments

We would like to thank many for helpful discussions related to the interpretation of this rich dataset, including Phil Appleton, Bruce Draine, Rosalie McGurk, Patrick Ogle, Rick Pogge, and Ralph Sutherland, and Emanuele Nardini and Junfeng Wang for sharing their processed X-ray maps.  We also thank the referee for careful readings and especially the scientific editor for patience while pandemic situations delayed our responses.

The authors wish to recognize and acknowledge the very significant cultural role and reverence that the summit of Maunakea has always had within the indigenous Hawai'ian community; we are privileged to be guests on your sacred mountain.  
We wish to pay respect to the Gamilaraay/Kamilaroi language group Elders - past, present and future - of the traditional lands on which Siding Spring Observatory stands, and to the Atacame\~{n}o community of the Chajnantor Plateau, whose traditional home now also includes the ALMA observatory.

Support for AMM was provided in part by NASA through Hubble Fellowship grant \#HST-HF2-51377 awarded by the Space Telescope Science Institute, which is operated by the Association of Universities for Research in Astronomy, Inc., for NASA, under contract NAS5-26555, and in part from the National Science Foundation under Grant No. 2009416.  AMM and LJK acknowledge the support of the Australian Research Council (ARC) through Discovery project DP130103925.  Parts of this research were supported by the Australian Research Council Centre of Excellence for All Sky Astrophysics in 3 Dimensions (ASTRO 3D), through project number CE170100013.  CEM acknowledges support by the National Science Foundation under award number AST-0908796.  KLL is supported by NASA through grants HST- GO-13690.002-A and HST-GO-15241.002-A from the Space Telescope Science Institute, which is operated by the Association of Universities for Research in Astronomy, Inc., under NASA contract NAS5-26555. 
T.D-S. acknowledges support from ALMA-CONICYT project 31130005 and FONDECYT regular project 1151239.
ET acknowledges support from CATA-Basal AFB-170002, FONDECYT Regular grant 1190818, ANID Anillo ACT172033 and Millennium Nucleus NCN19\_058 (TITANs).

This research is based on observations made with the NASA/ESA Hubble Space Telescope obtained from the Space Telescope Science Institute, which is operated by the Association of Universities for Research in Astronomy, Inc., under NASA contract NAS 5-26555. These observations are associated with programs 12552, 13690, and 10592.

Some of the data presented herein were obtained at the W. M. Keck Observatory, which is operated as a scientific partnership among the California Institute of Technology, the University of California and the National Aeronautics and Space Administration. The Observatory was made possible by the generous financial support of the W. M. Keck Foundation.  We enthusiastically thank the staff of the W. M. Keck Observatory and its AO team for their dedication and hard work.  

This paper makes use of the following ALMA data: ADS/JAO.ALMA\#2015.1.00370.S and \\
ADS/JAO.ALMA\#2015.1.00003.S. ALMA is a partnership of ESO (representing its member states), NSF (USA) and NINS (Japan), together with NRC (Canada) and NSC and ASIAA (Taiwan) and KASI (Republic of Korea), in cooperation with the Republic of Chile. The Joint ALMA Observatory is operated by ESO, AUI/NRAO and NAOJ.  The National Radio Astronomy Observatory is a facility of the National Science Foundation operated under cooperative agreement by Associated Universities, Inc.


\vspace{5mm}
\facilities{HST(ACS,WFC3),ATT(WiFeS),Keck:II(Laser Guide Star Adaptive Optics, OSIRIS, NIRC2), ALMA, CXO(ACIS)}

\software{LZIFU \citep[v1.1,][]{LZIFU}, PyWiFeS \citep[v0.6.0,][]{Childress14}
          }


\appendix

\section{Seeing-Limited Emission Line Maps from the WiFeS Dataset}
\label{wifesappendix}

Selected analysis products from our WiFeS integral field spectroscopy have been presented in the main body of the paper.  For completeness and for comparison with the higher spatial resolution emission line maps from narrowband imaging above, we present all remaining WiFeS emission line flux maps here, as well as some further analysis.

Using LZIFU \citep{LZIFU}, we fit up to three Gaussian components to each emission line; statistical F-tests show that all are justified (and perhaps even more) in most of the spaxels.  However, at these limited spatial resolutions, a galaxy as complex as NGC~6240 has many overlapping physical components that are smeared out: two nuclei as well as numerous filaments, bubbles, outflows, clusters, tidal tails and streams.  As such, we caution the reader against too much interpretation of these data and do not present kinematic maps of the individual components.  

\begin{figure}
\centering
\includegraphics[scale=1.,trim=.5cm 0.25cm 0cm 0.75cm]{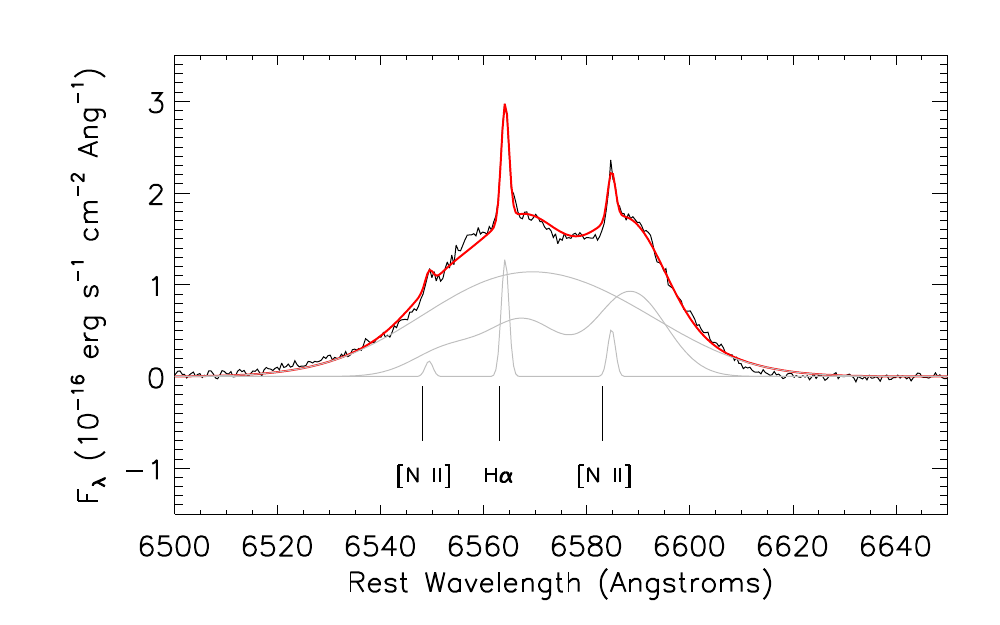}
\caption{Example continuum-subtracted red-arm spectrum from our WiFeS data of a single 1\arcsec $\times$ 1\arcsec~spaxel near the nuclei, zoomed in on the \ha-\nii~complex to demonstrate the challenging kinematic blending.  We show in grey the LZIFU-fit three individual kinematic components for each line, and the sum of all line components in red.}
\label{wifesblending}
\end{figure}

\begin{figure*}
\centering
\includegraphics[scale=.9,trim=1cm 1cm 0cm 1cm]{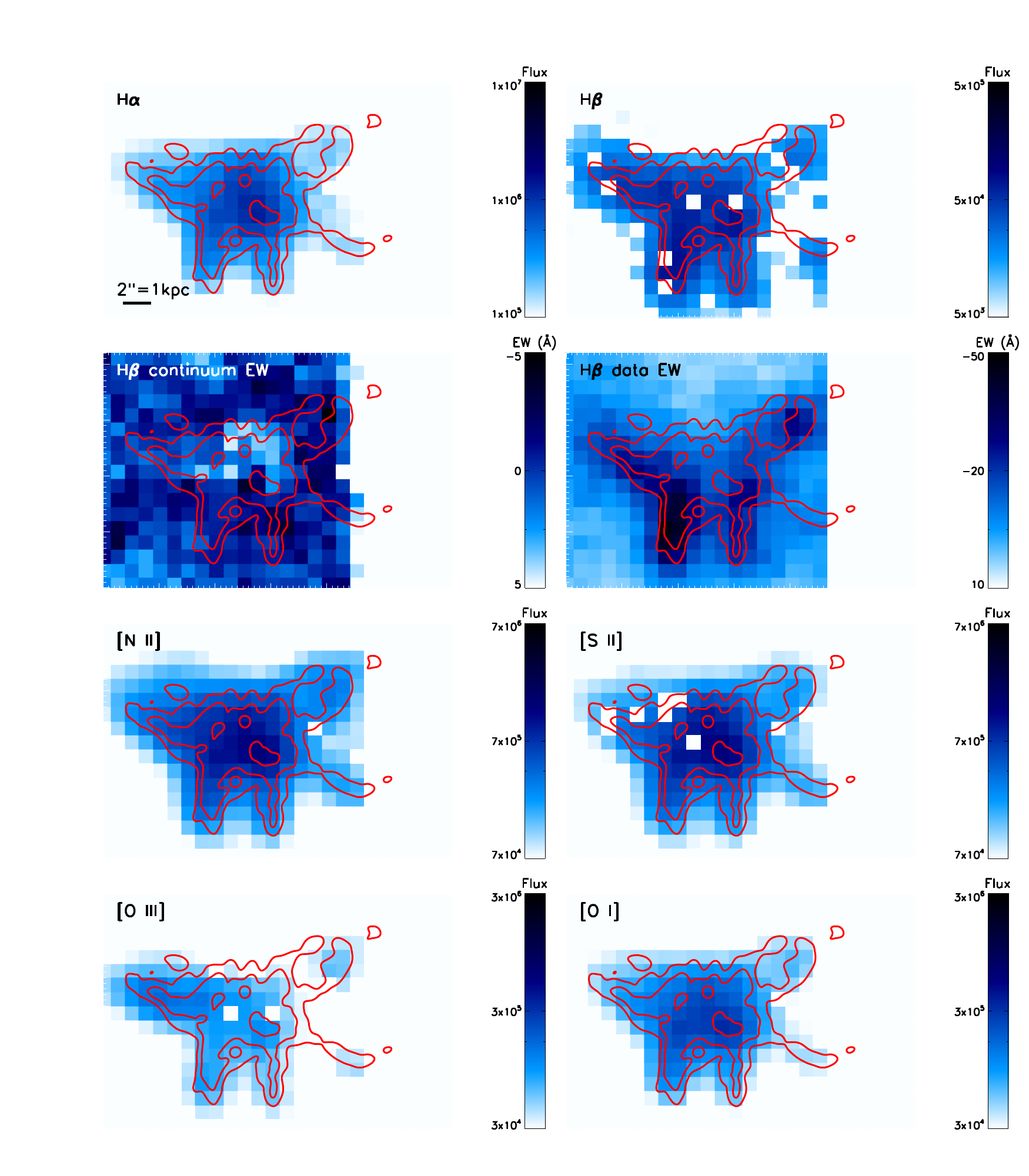}
\caption{Maps of emission lines \ha, \hb, \nii, \sii, \oiii, and \oi~from WiFeS seeing-limited integral field spectroscopy.  Each panel is shown in log scale with limits chosen to emphasize structure, and has the HST \ha+\nii~contours (red) overlaid for comparison.   Line emission flux maps are shown in units of 10$^{-20}$ erg s$^{-1}$ cm$^{-2}$ arcsec$^{-2}$.  The second row shows the equivalent widths of \hb in the continuum fit (left) and for the overall datacube (right), to illustrate the effects of Balmer absorption on the HST narrowband images: up to 30\% in the nuclei and minor in the outer arms.  A Negative EW (darker colors) represents emission and positive EW (lighter colors) absorption.  The scale bar in the top left panel shows 2.05 arcseconds, approximately one kiloparsec.  Spaxels for which the emission line fit failed or for which the S/N$<$3 are masked in white.
}  
\label{wifeslinemaps}
\end{figure*}

In Figure~\ref{wifeslinemaps}, we show the emission line flux maps.  Because many of the spectral profiles are so complex that even three Gaussian components may not accurately describe them, we hesitate to attribute pure physical meaning to any individual component.  Therefore, we show the sums of the three Gaussians for each line for each spaxel.  Even so, we favor the parametrized framework in order to do the best job deblending the \ha~and \nii~lines.  We note that the high spectral resolution of the WiFeS red arm (R$\sim$7000) is capable of deblending the \ha~and \nii~lines easily in a normal system, but the complex kinematics and broad components of the lines pose a challenge that spectral resolution cannot solve (Figure~\ref{wifesblending}).  IFS with improved spatial resolution may provide some opportunity to spatially separate the complex kinematic structures into more distinct line profiles in the future.

We also show the key diagnostic line ratio maps \nii/\ha, \sii/\ha, \oi/\ha, and \oiii/\hb in Figure~\ref{wifes_lineratios} following Figure~\ref{ratiomaps}.  Although these maps are at significantly lower spatial resolution than those constructed from the HST narrowband images, they have an important feature: we are able to spectrally separate \ha~and \nii, so the line ratios are purely forbidden transitions to recombination lines.  The HST images constructed from \sii/(\ha$+$\nii) and \oi/(\ha$+$\nii) show much less enhancement along the filaments than the pure line ratios \sii/\ha~and \oi/\ha~probed by our WiFeS data.  This effect is due to a cospatial enhancement of \nii~along the same filaments.  The \oiii/\hb~map also shows an enhanced line ratio along the eastern arm, but the lower spatial resolution washes it out considerably.  The most surprising feature revealed from the WiFeS data is the variation between the \nii/\ha~and \sii/\ha~ratio maps; in particular, \nii~is enhanced to the northeast and northwest of the Butterfly Nebula.  Indeed, diffuse \nii~and \ha~extend in our WiFeS data $\simgt$5 kpc north of the upper edge of the panels in Figures~\ref{wifeslinemaps} and~\ref{wifes_lineratios}, as was also seen in the narrowband images of \citet{Veilleux03}.  This diffuse emission is detected at low significance in our HST F673N and F680N images; the detections are below our 3$\sigma$ cutoff and are therefore masked out.  However, extended diffuse \sii~is either considerably weaker or undetected even in our WiFeS dataset.

\begin{figure*}
\centering
\includegraphics[scale=.9,trim=1cm 12.5cm 0cm 1cm]{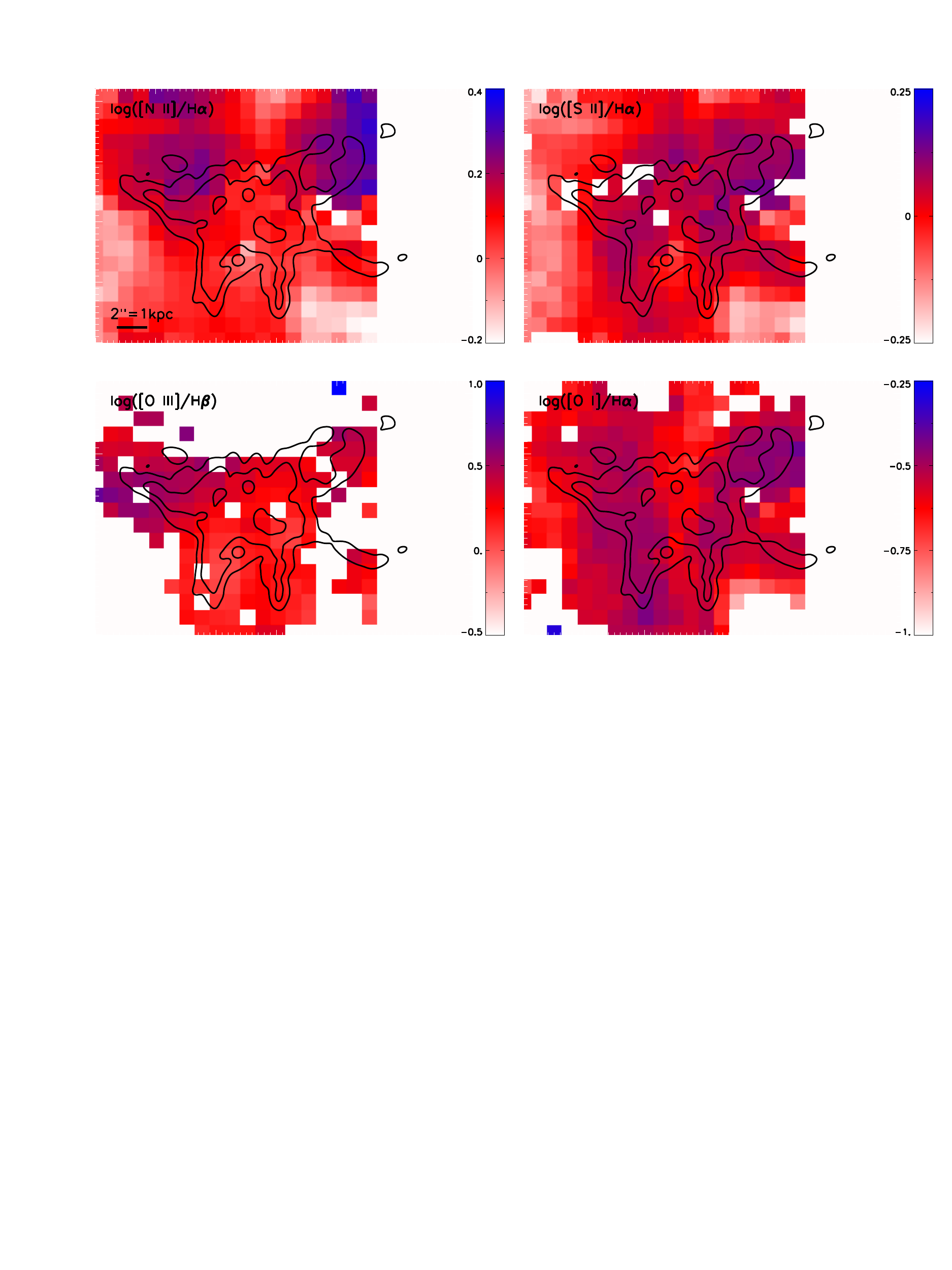}
\caption{Emission line ratio maps \nii/\ha, \sii/\ha, \oi/\ha, and \oiii/\hb~constructed from WiFeS seeing-limited integral field spectroscopy.  By spectrally separating \ha~and \nii, we see an enhancement of \sii/\ha~and \oi/\ha~along the filaments that was not evident in the HST dataset shown in Figure~\ref{ratiomaps}.  
}  
\label{wifes_lineratios}
\end{figure*}

\begin{figure*}
\centering
\includegraphics[scale=1.2,trim=1cm 1.11cm 0cm 0.5cm]{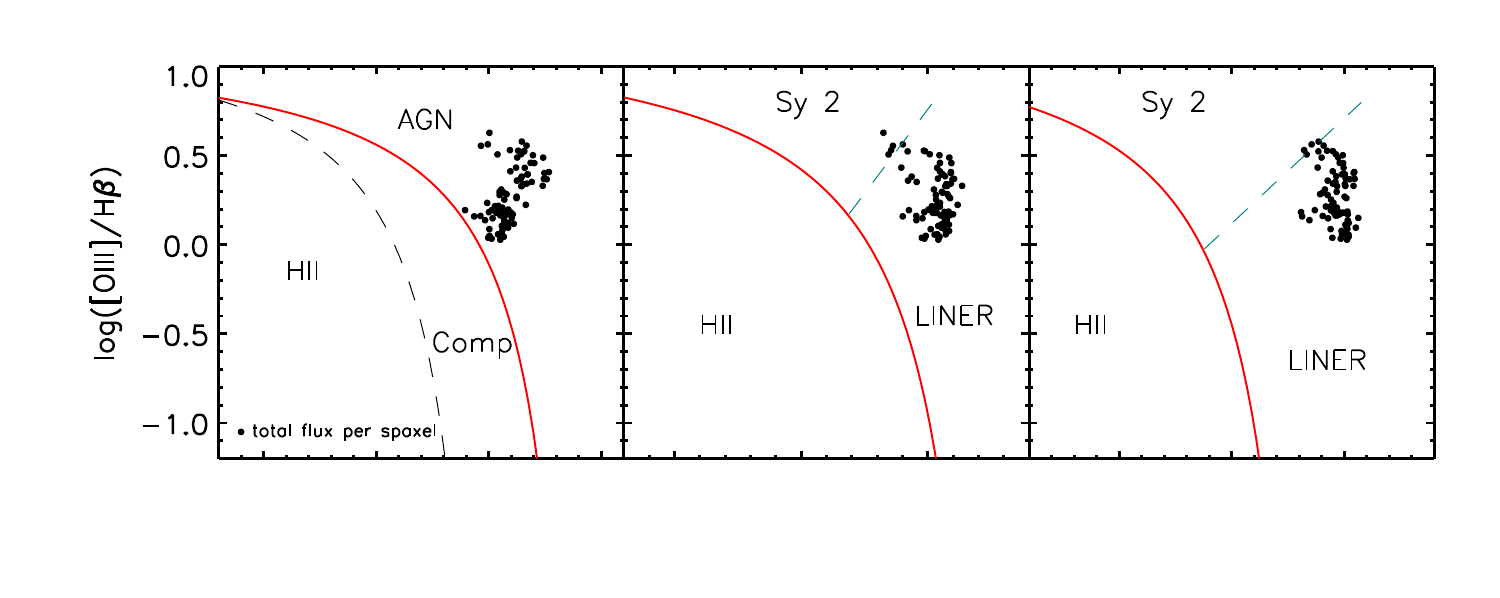}
\caption{Emission line diagnostic diagrams from \citet{BPT} and \citet{Veilleux87} and classification lines from \citet{Kewley06} for each spaxel of our WiFeS integral field spectroscopy data.  In each panel, each point shows the summed flux from all Gaussian components fit to each emission line.  Data are only shown if all relevant emission lines have a S/N$>$5.  As expected from previous unresolved spectra, most of NGC~6240's emission is dominated by LINER-like line ratios, most likely due to shocks caused by the ongoing outflows.  The diagnostic diagrams show a shock mixing sequence (extending from bottom left towards top right of each panel) with an upturn towards the top left where the high \oiii\hb~ratio suggests AGN-like photoionization.
}  
\label{wifesBPT}
\end{figure*}

In Figure~\ref{wifesBPT}, we translate the line ratio maps from Figure~\ref{wifes_lineratios} into the the optical emission line diagnostic diagrams from \citet{BPT} and \citet{Veilleux87}.  The spectra across the field-of-view show elevated line ratios mostly falling in the LINER classification region from \citet{Kewley06}.  We see evidence of a shock-mixing sequence extending from lower line ratios to higher in all three panels, likely demonstrating spatial regions where ionization from outflow-induced shocks dominates more (top right) or less (bottom left) over star formation \citep{Rich11,Davies17,DAgostino18}.  A small turn to the upper left is caused by the high \oiii/\hb~ratio seen in the eastern arm discussed in Section~\ref{oiiiarm}.

\bibliography{bibGOALS,bibWiFeS,bibH2,bibNGC6240}
\bibliographystyle{aasjournal}

\end{document}